\documentclass[preprint,12pt]{elsarticle}

\usepackage{iftex}
\usepackage[T1]{fontenc}
\usepackage[utf8]{inputenc}
\ifPDFTeX
\DeclareUnicodeCharacter{00A7}{\S}
\DeclareUnicodeCharacter{00B1}{\ensuremath{\pm}}
\DeclareUnicodeCharacter{00B7}{\ensuremath{\cdot}}
\DeclareUnicodeCharacter{00D7}{\ensuremath{\times}}
\DeclareUnicodeCharacter{03B1}{\ensuremath{\alpha}}
\DeclareUnicodeCharacter{03BB}{\ensuremath{\lambda}}
\DeclareUnicodeCharacter{03C3}{\ensuremath{\sigma}}
\DeclareUnicodeCharacter{03C4}{\ensuremath{\tau}}
\DeclareUnicodeCharacter{0393}{\ensuremath{\Gamma}}
\DeclareUnicodeCharacter{2013}{--}
\DeclareUnicodeCharacter{2014}{---}
\DeclareUnicodeCharacter{2018}{`}
\DeclareUnicodeCharacter{2019}{'}
\DeclareUnicodeCharacter{201C}{``}
\DeclareUnicodeCharacter{201D}{''}
\DeclareUnicodeCharacter{2026}{\ldots}
\DeclareUnicodeCharacter{2032}{\ensuremath{'}}
\DeclareUnicodeCharacter{2074}{\textsuperscript{4}}
\DeclareUnicodeCharacter{2077}{\textsuperscript{7}}
\DeclareUnicodeCharacter{207B}{\textsuperscript{--}}
\DeclareUnicodeCharacter{2190}{\ensuremath{\leftarrow}}
\DeclareUnicodeCharacter{2192}{\ensuremath{\rightarrow}}
\DeclareUnicodeCharacter{21D2}{\ensuremath{\Rightarrow}}
\DeclareUnicodeCharacter{2205}{\ensuremath{\emptyset}}
\DeclareUnicodeCharacter{2208}{\ensuremath{\in}}
\DeclareUnicodeCharacter{2227}{\ensuremath{\wedge}}
\DeclareUnicodeCharacter{222A}{\ensuremath{\cup}}
\DeclareUnicodeCharacter{2212}{\ensuremath{-}}
\DeclareUnicodeCharacter{2248}{\ensuremath{\approx}}
\DeclareUnicodeCharacter{2264}{\ensuremath{\leq}}
\DeclareUnicodeCharacter{2265}{\ensuremath{\geq}}
\DeclareUnicodeCharacter{227C}{\ensuremath{\preccurlyeq}}
\DeclareUnicodeCharacter{2286}{\ensuremath{\subseteq}}
\DeclareUnicodeCharacter{2287}{\ensuremath{\supseteq}}
\DeclareUnicodeCharacter{2295}{\ensuremath{\oplus}}
\DeclareUnicodeCharacter{22A2}{\ensuremath{\vdash}}
\DeclareUnicodeCharacter{1D54B}{\ensuremath{\mathbb{T}}}
\DeclareUnicodeCharacter{2713}{\ensuremath{\checkmark}}
\DeclareUnicodeCharacter{2717}{\ensuremath{\times}}
\fi
\usepackage{lmodern}
\usepackage{amsmath,amssymb}
\usepackage{booktabs}
\usepackage{array}
\usepackage{graphicx}
\usepackage{caption}
\usepackage{algorithm}
\usepackage{algpseudocode}
\usepackage{listings}
\usepackage{tikz}
\usetikzlibrary{arrows.meta,positioning,fit,calc,patterns,patterns.meta}
\usepackage{pgfplots}
\usepgfplotslibrary{groupplots}
\pgfplotsset{compat=1.18}
\usepackage{url}
\usepackage[hidelinks]{hyperref}
\usepackage{microtype}
\usepackage{placeins}
\journal{Data \& Knowledge Engineering}
\biboptions{sort&compress}

\begin{document}

\begin{frontmatter}

\title{VirtualSet: Receiver-Indexed Authority Compilation for Query-to-Action Continuity in Executable Ontologies}

\author[sjtu]{Qunhui Zhang\corref{cor1}}
\ead{will_zhang@sjtu.edu.cn}

\cortext[cor1]{Corresponding author}
\address[sjtu]{School of Software, Shanghai Jiao Tong University,
800 Dongchuan Road, Shanghai 200240, China}

\begin{abstract}
An ontology edit can preserve a query answer while replacing the evidence that justified an
action on it. We study when the original approval remains valid. VirtualSet compiles the
evaluated query and a trusted grant into a single-use capability. The query result fixes
the approved objects; policy grants actions on their type. Each field or link is recorded
under its owning receiver type. For conjunctive filters over stable,
single-valued relations, preserving the observed paths and target rows suffices for current
target membership at batch entry. In 32 Java cases requiring recorded evidence, finite paths
admitted 24 continuations versus 16 for a type-level closure. We then issued approvals before
controlled changes in procurement and laboratory ontologies and presented the original
credentials afterward. In all 40 scenarios, the post-change selections returned the same
object identifiers as at issuance. With the same Java authorization layer, typed navigation
and SQL over a stable view each retained 22 original approvals and blocked 18. Retargeting
the second relation invalidated approvals that traversed it while preserving those based
only on the first. By tracing the receiver type at each access in a validated query, VirtualSet
distinguishes approvals for the same objects according to the evidence each approval depends on.
\end{abstract}


\end{frontmatter}
\hypersetup{pdfauthor={Qunhui Zhang}}

\section{Introduction}
\label{sec:introduction}

Ontologies and conceptual models give data-intensive systems a stable vocabulary for entities,
relations, constraints, and operations above heterogeneous stores
\cite{gruber-ontology-1993,poggi-obda-2008,calvanese-ontop-2017}. Ontology-based data access has
concentrated on the read side of this vocabulary: conceptual queries are rewritten to sources and
their answers are lifted back to the conceptual level. Operational knowledge systems increasingly
also expose the \emph{write} side: an ontology names the actions that may be applied to its
concepts. A program generated by a large language model (LLM), for example, may select objects
through an ontology query and later invoke an authorized action on those objects.
The resulting program must carry the ontology's meaning through intermediate computations and
into the operation that consumes their result.

This setting requires continuity at two boundaries. During query composition, the result of
each operation must expose the appropriate ontology members. For example, filtering a
\texttt{Countries} collection should preserve its domain-specific \texttt{provinces()}
navigation, whereas projecting country names should establish a string-valued receiver.
Receiver-sensitive typing provides this structural continuity.

The second boundary separates a typed result from a later action. Text-to-SQL systems
produce a query and its answer \cite{li-bird-2023,scholak-picard-2021}; tool interfaces
describe operations and arguments \cite{li-api-bank-2023,yao-taubench-2025}.
A correct query answer and an applicable operation establish which objects qualify and
which action their type supports. Neither establishes whether an approval issued for
those objects remains valid when the ontology or the underlying data change.

Consider a purchase order selected for approval on the strength of its supplier's certificate.
When the supplier relation for certificates is retargeted to a new entity type, a join to
the retired certificate table can still return the correct purchase ID. What has changed
is the source of the evidence, not the identity of the object awaiting approval. Reading
the old source can have either consequence: approval can survive a change to the current
certificate, or lapse after a change to a certificate no longer used. A delayed approval
must identify both the objects that may receive the action and the observations that justify it.

We model an approval as a contract for one evaluation. A \emph{query-derived receiver
capability} (QRC) binds the selected identifiers $D$ and the evidence for that selection
to a trusted grant specifying the action and its arguments. A later invocation names a
nonempty subset $A\subseteq D$. If the action requires the original observations, the
dispatcher checks the recorded values and links for $D$ even when $A$ has only one member.

VirtualSet constructs this contract by \emph{receiver-indexed authority compilation}.
Generic Constraints Projection (GCP) validates the query against the current
ontology~\cite{dli-paper}. A separate analysis traverses the validated expression,
tracking the receiver at each access and looking up the field or link on that receiver.
This identifies the owner of each observation; the result type identifies the receiver
on which the controller may grant an action. The distinction
matters when two types declare the same member. Checking the root's field can miss a
change on a linked object; checking every field with that name can invalidate an approval
after an unrelated change. The Java attribution experiment isolates these two errors
(Table~\ref{tab:cross-stage-ablation}).

Action descriptors specify the required continuity level: historical selection (G1) preserves
approved targets and their issue-time row state; current target membership (G2) additionally
requires the targets to satisfy the selecting predicate when the batch first enters action code.
The Java descriptor independently specifies whether the action also requires equality of
the recorded observations. Issuance, dispatch, and action entry enforce both requirements.
If a supplier's rating changes from 80 to 90 against a threshold of 70, rechecking establishes
that the order still qualifies; a policy tied to the recorded rating requires renewed approval.

The contributions are as follows.

\begin{itemize}
\item \textbf{A contract for existing approvals.} The contract distinguishes the objects
selected at issuance, the evidence an action requires to remain current, and the authority
granted by policy. Theorems~1 and~3 relate those three elements to the decision at action
entry. They apply to the represented dependencies and the trusted version providers.
\item \textbf{Receiver-indexed evidence derivation.} The compiler follows the receiver
type through the query, including changes of owner inside predicates, while retaining
the result type for action lookup. Lemma~1 establishes coverage of referenced members.
For row-local functional filters, Theorem~2 shows when the captured evidence suffices
to establish target membership without repeating the selection.
\item \textbf{Approval validity under ontology evolution.} Two selections can return
the same objects under the same grant yet depend on different evidence. In a
\hyperref[exp:java-equal-answers]{paired Java experiment}, changing one endpoint
invalidates only the approval that used it. A separate SQLite comparison isolates the
effect of binding the selection to the current or retired evidence source
(Table~\ref{tab:owner-intervention}). Presenting credentials issued before a relation
was retargeted shows which approvals survive the change: those whose recorded path
uses the retargeted relation are blocked even when their answer is unchanged
(Table~\ref{tab:old-approval}).
\end{itemize}

Sections~\ref{sec:related} and~\ref{sec:runtime} position the problem and introduce the
ontology runtime. Sections~\ref{sec:model} and~\ref{sec:execution} develop the contract,
its properties and its implementation; Sections~\ref{sec:evaluation}--\ref{sec:conclusion}
report the evaluation and its implications.

\section{Related Work}
\label{sec:related}

\subsection{Ontology-mediated data and executable knowledge interfaces}

Ontology-based data access (OBDA) connects a conceptual model to relational sources through
mappings and query rewriting \cite{poggi-obda-2008,calvanese-ontop-2017}. SPARQL, Cypher,
Gremlin, and TypeQL provide semantic or graph-oriented navigation, while LookML, Malloy, Cube, and
dbt curate domain concepts above storage schemas
\cite{w3c-sparql-2013,opencypher-2026,tinkerpop-gremlin-2026,typeql-2026,
google-lookml-2026,malloy-docs-2026,cube-docs-2026,dbt-docs-2026}. Ontology-to-tool compilation
and commercial ontology platforms extend the semantic layer toward typed objects, policies, and
actions \cite{zhou-ontology-tools-2026,palantir-ontology-2026,palantir-actions-2026}.
In particular, ontology-to-tools compilation derives executable interfaces for creating and
modifying knowledge-graph instances while enforcing semantic constraints during generation
\cite{zhou-ontology-tools-2026}. QRC compiles a different object: authority for the concrete
denotation of an already evaluated query. Its dependency analysis determines which ontology
and data changes preserve that authority until a later mutation. The two approaches thus
operate at complementary schema-to-interface and evaluated-query-to-authority boundaries.

Update management in OBDA studies how ontology-level insertions and deletions translate to
source updates while preserving mappings and constraints \cite{degiacomo-obda-updates-2017}.
Query inseparability compares knowledge bases by the answers they give to a class
of queries over a specified vocabulary \cite{botoeva-inseparability-2016}. Approval
validity concerns a grant issued for one particular evaluation. An edit can preserve the
selected identifiers yet change the source of a recorded observation; that distinction
matters when the original approval is presented for action.

\subsection{Typed read--write interfaces and receiver preservation}

Receiver preservation belongs to an established line of language research. F-bounded
quantification and self types express relationships between a receiver and its method signatures
\cite{canning-fbound-1989,bruce-binary-1995}. Self type constructors explicitly address recursive
generic interfaces whose result remains in the same collection family while its element type
changes \cite{saito-self-2009}. Java's \texttt{Stream.map} returns
\texttt{Stream<R>}; TypeScript's polymorphic \texttt{this} supports subtype-sensitive fluent
results; Scala's collection framework provides construction protocols for collection-specific
results \cite{java-stream-2026,typescript-this-2026,scala-collections-2026}. Element-type inference,
self types, and collection-family reconstruction are therefore distinct comparison dimensions.

GCP provides call-site projection, Outline Equational Matching, and future-\texttt{this}
\cite{dli-paper}. VirtualSet uses these established rules to obtain a receiver for each
ontology expression. At an access to a field or relation, VirtualSet uses that receiver
to identify the declaration and the observations on which an approval depends. At the
result, it binds the selected identifiers to a trusted grant. At dispatch, the
runtime checks the targets and grant and enforces the continuity and evidence
requirements declared by the action.

GraphQL defines typed query and mutation roots, with top-level mutation fields executed serially
\cite{graphql-spec-2021}; LINQ supplies fluent typed query construction within a host language
\cite{meijer-linq-2006}. OData is the closest protocol-level predecessor to receiver-bound
execution: its CSDL distinguishes side-effect-free functions from actions and binds both to entity
or collection resources \cite{odata-csdl-2020}. A resource path can invoke an applicable bound
operation after filtering, and \texttt{\$each} applies an action to individual members of a
collection \cite{odata-url-2020}. OData also supports optimistic concurrency through entity
tags: a client may condition a later request, including a bound action request, on an
\texttt{If-Match} header carrying the ETag of the addressed entity or collection, and the service
rejects the request when that resource has changed
\cite{odata-protocol-2020}. This is the closest protocol-level predecessor of QRC's row-state
fingerprints: the service supplies the resource tag, and the client presents it as a
precondition on the later request.

QRC binds an operation to an evaluated query result and its receiver-qualified dependencies,
combining target-row freshness with referenced ontology and policy versions in one authority
object. An application can obtain equivalent continuation decisions using a saved result,
row snapshots, and transactional SQL rechecking. QRC's contribution is the derivation of
that contract from the ontology expression and its enforcement against each action's
declared requirement. Section~\ref{sec:northwind} compares these evidence paths with an
independent JDBC controller; Table~\ref{tab:authority-position} summarizes their neighboring
interface and authorization mechanisms.

\subsection{Capabilities and fine-grained authorization}
\label{sec:related-capabilities}

Capability systems represent authority as an unforgeable reference whose possession and content
determine the permitted operation \cite{dennis-vanhorn-1966,miller-robust-composition-2006}.
Contextual credentials such as macaroons add caveats that
restrict who may exercise authority, for which operation, and under which conditions
\cite{birgisson-macaroons-2014}; Biscuit tokens carry Datalog facts and checks that are
attenuated offline and evaluated against verifier-side facts \cite{biscuit-2026}. Rich
Authorization Requests encode fine-grained resource
and action details and require integrity protection against tampering or swapping
\cite{rfc9396-rar}. These mechanisms motivate authenticated attenuation, but their authorization
details are supplied by a client or authorization service rather than derived from an evaluated
query. Usage control extends access control with ongoing authorization: a decision made at
request time is re-evaluated against mutable attributes while access continues
\cite{park-sandhu-ucon-2004}. QRC's dependency premise is an instance of ongoing authorization
in which the mutable attributes are the fingerprints of the ontology members, policies, and rows
that justified the original decision.

Provenance semirings describe how input tuples contribute to a query result
\cite{green-provenance-2007}. M\"uller et al.\ rewrite SQL to derive cell-level dependencies
that explain where result values originate and why they appear \cite{muller-sql-provenance-2018}.
Such rewriting identifies the cells read by the SQL that actually ran. If a join reads a
retired inventory table, attributing the answer to that table is correct SQL provenance.
Whether the table still implements the relation named in an approval is a separate
mapping judgment. VirtualSet resolves the owner of each read along the navigation path
and looks up action descriptors on the result's entity type. The selected objects constrain the targets and
trusted grants constrain the operation. In Section~\ref{sec:prototype} a
restricted SQL extractor uses join aliases, keys, and predicate columns instead of typed paths.
We supply it with the same object mapping and capability layer and test the resulting actions.
This experiment compares automatic derivations for their shared query fragment.
Section~\ref{sec:contract-evolution} separately measures the effect of editing an explicit
JDBC observation projection.

Incremental view maintenance addresses a different question: how to update an answer as the
database changes. DBToaster derives higher-order delta computations for this purpose
\cite{ahmad-dbtoaster-2012}. A maintained answer can show that an object still qualifies;
whether a change to the evidence requires renewed approval is a policy decision.

Database authorization has long used views and query modification to constrain the tuples visible
to a principal, and query rewriting has been extended to fine-grained, cell-level access control
\cite{stonebraker-query-modification-1974,griffiths-wade-1976,rizvi-fgac-2004}. Provenance-aware
authorization relates access decisions to the dependency paths that produced or justify data
objects \cite{nguyen-provenance-access-2012}. In that setting a dependency path is an
input to an access policy. Here the selection determines which observations to bind;
a separate trusted policy determines the permitted update. The two judgments meet at
action entry, when a principal proposes to act on a subset of the earlier answer. This
resembles optimistic read-set validation \cite{kung-robinson-occ-1981}, but spans two
requests, and the action descriptor determines the required guarantee.

ShillDB's CapQL interface derives view capabilities by filtering, projecting or joining
views; contracts restrict the operations permitted through each capability
\cite{zigmond-shilldb-2020}. The capability represents a view expression, which CapQL
evaluates when a database operation is invoked. QRC instead binds a grant to the objects
returned by a particular evaluation. If the action requires unchanged evidence, dispatch
also checks the recorded fields and links, even when the same objects still qualify.
Table~\ref{tab:authority-position} compares CapQL and QRC at the level of their
authorization semantics.

The evaluation isolates these authorization dimensions with two contract ablations. The
signed-resource condition authenticates principal, identifiers, action, arguments, effect,
and nonce. The resource-version condition additionally binds target rows. Comparing both
with QRC measures the contribution of query provenance and receiver-qualified ontology,
action, policy, and executor dependencies.

A carried, self-describing credential is one of two placements for such authority. The other is
a \emph{server-side result handle}: the runtime keeps the authorization state---result set,
object versions, permitted operations, one-time token---and hands the client an opaque reference,
as opaque OAuth tokens do with introspection at the authorization server \cite{rfc7662-2015} and
as session-bound cursors do for query results. The placement question is orthogonal to what the
authority contains; a handle design can store the same query-derived footprint that a QRC
carries. Section~\ref{sec:evaluation} therefore also implements a result-handle design over the
same derived payload and decision rule. Holding both fixed isolates the effect of placement
on dispatch decisions and cost.

Table~\ref{tab:authority-position} distinguishes interface-level guarantees from the
application-defined continuation contract. The signed-resource, resource-version, and
global-version rows are executable contract ablations. The automatic SQL row reports
the combined extractor and authorization layer evaluated in Section~\ref{sec:prototype};
the SQL provenance row describes rewriting alone.

\begin{table}[!htbp]
\centering
\caption{Authority continuity across mechanisms}
\label{tab:authority-position}
\scriptsize
\setlength{\tabcolsep}{2.7pt}
\begin{tabular}{>{\raggedright\arraybackslash}p{2.85cm}>{\raggedright\arraybackslash}p{2.25cm}>{\centering\arraybackslash}p{1.65cm}>{\centering\arraybackslash}p{1.65cm}>{\centering\arraybackslash}p{1.90cm}>{\centering\arraybackslash}p{1.35cm}}
\toprule
Mechanism & Target source & Query provenance & Exact target binding & Selective freshness & At-most-once \\
\midrule
OBDA query rewriting & Rewritten conceptual query & -- & -- (read only) & -- & -- \\
SQL provenance rewriting & SQL result cells & Cell lineage & External & External & External \\
GraphQL mutation & Mutation arguments & -- & External & External & External \\
OData bound operation & Resource path & -- & Path-level & External & External \\
Receiver-scoped catalog (Sec.~\ref{sec:runtime}) & Receiver type & -- & Type-level & -- & -- \\
Macaroon / Biscuit / RAR & Credential policy & Optional caveat & Optional & Caveat-level & External \\
OData conditional operation & Entity/collection path & -- & Resource-level & Resource ETag & External \\
ShillDB/CapQL (semantic comparison only) & Derived view & View expression & Current view rows & No issue-time evidence check & External \\
CaMeL value capability & Interpreter data flow & Data-flow provenance & Policy-level & -- & -- \\
Signed-resource ablation & Authenticated identifiers & -- & Yes & -- & Yes \\
Resource-version ablation & Authenticated identifiers & -- & Yes & Target row versions & Yes \\
Global-version ablation & Evaluated query result & Yes & Yes & Whole snapshot & Yes \\
Automatic SQL extractor + shared authority (Sec.~\ref{sec:prototype}) & SQL-selected objects & Signed SQL digest & Yes & Target rows, join keys, predicate fields & Shared ledger \\
QRC & Evaluated ontology-query result & Yes & Yes & Referenced knowledge and state & Yes \\
\bottomrule
\end{tabular}
\end{table}

\subsection{LLM program generation and operational tools}

Text-to-SQL systems produce a relational program and its answer; constrained decoding,
schema linking, and repair improve that translation
\cite{yu-spider-2018,li-bird-2023,scholak-picard-2021,pourreza-2023-dinsql,wang-mac-sql,
gao-dailsql-2024,gao-xiyansql-2024,eckmann-hlr-sql-2026,bird-interact-2026}. Other systems
constrain or validate the generated program against a language or type
\cite{poesia-synchromesh-2022,beurerkellner-2023-lmql,willard-2023,microsoft-typechat-2023}.
Neither a well-typed program nor a correct answer alone specifies who may subsequently act
on which objects.

Stateful tool benchmarks test action selection, execution, and recovery
\cite{li-api-bank-2023,ruan-toolemu-2024,yao-taubench-2025,lu-toolsandbox-2025,
debenedetti-agentdojo-2024}. CaMeL attaches capabilities to program values, Progent monitors
tool-call privilege, and AuthGraph compares intended authority with execution provenance
\cite{debenedetti-camel-2025,shi-progent-2025,wang-authgraph-2026}. PACE checks the
effects of a proposed call against authority derived from an authenticated request,
using provenance to constrain its inputs~\cite{li-pace-2026}. SecureGARD uses ontology
rules and task state to determine whether a proposed call can proceed, must wait, or
is forbidden~\cite{xiong-securegard-2026}. QRC addresses a decision spanning two
requests: whether the evidence and grant bound to an earlier query result still justify
acting on it. The dispatcher checks this at local action entry; subsequent delivery of external effects is a separate transaction
problem \cite{saltzer-schroeder-1975,sandhu-rbac-1996,gray-transactions-1981,garcia-molina-sagas-1987}.

\section{Ontology Runtime and Problem Statement}
\label{sec:runtime}

\subsection{System architecture}

VirtualSet's base runtime is organized as a four-stage pipeline: runtime context assembly,
model-facing construction, receiver validation, and execution. The System Outline defines the shared
VirtualSet algebra; the current World Outline contributes entities, edges, and actions; and the
action registry supplies the descriptors applicable to those receivers. For QRC issuance, the
authority layer captures a revisioned stable view of the referenced descriptors. Global and world
recipes provide generation guidance, but the two Outlines and the registry determine structural
legality. The construction loop combines this context with the request, evidence, and sampled
values to produce a complete expression for GCP validation.

Fig.~\ref{fig:architecture} shows the execution split after validation. Translatable reads use
\texttt{GCPToSQL}, and unsupported read tails use the interpreter-backed
\texttt{StreamSetValue}; the principal-scoped \texttt{queryIds} boundary performs read
authorization and delegates typed identifier extraction to \texttt{evalFilter}.
Receiver-bound actions are resolved from
the \texttt{ActionRegistry} and retain the permission, descriptor, entity-loading, and event
semantics of \texttt{executeAction}. For a selection that is to justify a later action, the QRC
extension admits a translatable fragment that retains a root-entity result, records its dependency footprint,
and wraps the evaluated identifiers as denotation-bounded authority. The dispatcher validates
that authority before entering the existing receiver-bound mutation path. Built-in CRUD operations
publish typed events (\texttt{EntitirEvent}; Entitir is the host platform in which VirtualSet is
implemented); a custom action enters the event path when it performs an event-producing mutation
or publishes an event.

\begin{figure}[!htbp]
\centering
\resizebox{\linewidth}{!}{%
\begin{tikzpicture}[
  font=\footnotesize,
  box/.style={draw, rounded corners=2pt, align=center, inner sep=3.5pt},
  outline/.style={box, fill=gray!10, minimum width=2.7cm, minimum height=0.78cm},
  recipe/.style={box, fill=gray!3, minimum width=2.7cm, minimum height=0.78cm},
  stage/.style={box, fill=gray!8, minimum width=6.2cm, minimum height=0.82cm},
  engine/.style={box, fill=gray!12, text width=4.7cm, minimum height=0.82cm},
  eventbox/.style={box, fill=gray!14, text width=4.7cm, minimum height=0.82cm},
  frame/.style={draw=gray!70, densely dashed, thick, rounded corners=4pt, inner sep=8pt},
  arr/.style={-{Latex[length=2.0mm]}, thick},
  soft/.style={-{Latex[length=1.8mm]}, draw=gray!65, thick},
  deny/.style={-{Latex[length=1.8mm]}, dashed, draw=gray!65, thick}
]
\node[outline] (sys) at (-1.6,0) {\textbf{System Outline}\\VirtualSet algebra};
\node[recipe] (global) at (1.9,0) {\textbf{Global Recipe}\\shared guidance};
\node[outline] (wout) at (-1.6,-1.2) {\textbf{World Outline}\\entities, edges, actions};
\node[recipe] (wrec) at (1.9,-1.2) {\textbf{World Recipe}\\local conventions};
\node[frame, fit=(sys)(global)(wout)(wrec)] (ctx) {};
\node[font=\scriptsize, text=black!75, above=2pt of ctx.north]
  {LLM context: outlines (authority) + recipes (guidance)};
\node[font=\scriptsize] (req) at (0,-2.55) {request + evidence + values};
\node[stage] (aip) at (0,-3.45) {\textbf{construction loop}\\complete VirtualSet expression};
\node[stage] (gcp) at (0,-4.7) {\textbf{GCP + future-this}\\validated expression};
\node[engine] (sql) at (-2.55,-6.35) {\textbf{read branch}\\\texttt{queryIds}/\texttt{evalFilter}\\\texttt{GCPToSQL}/\texttt{StreamSetValue}};
\node[engine] (act) at (2.55,-6.35) {\textbf{QRC authority layer}\\receiver-flow analysis\\and grant verification};
\node[engine] (readrepo) at (-2.55,-8.15) {\textbf{ontology repository}\\read access};
\node[engine] (mutrepo) at (2.55,-8.15) {\textbf{receiver-bound action}\\registry + local mutation};
\node[engine, fill=gray!6] (result) at (-2.55,-10.0) {\textbf{entity identifiers} $D$\\mint authenticated QRC};
\node[eventbox] (event) at (2.55,-10.0) {\textbf{typed EntitirEvent}\\event-producing paths};
\node[engine, fill=gray!6] (real) at (2.55,-11.85) {\textbf{external integration}\\configured listener};
\draw[soft] (ctx.south) -- (req);
\draw[soft] (req) -- (aip);
\draw[arr] (aip) -- (gcp);
\draw[soft] (gcp.east) -- ++(1.1,0) |- node[pos=0.22, right, font=\scriptsize, align=left]
  {diagnostic +\\receiver closure} (aip.east);
\draw[arr] (gcp.south west) -- (sql.north);
\draw[arr] (gcp.south east) -- (act.north);
\draw[arr] (sql.south) -- node[left, font=\scriptsize] {read} (readrepo.north);
\draw[arr] (act.south) -- node[right, font=\scriptsize] {mutate} (mutrepo.north);
\draw[arr] (readrepo.south) -- (result.north);
\draw[arr] (result.east) -- ++(0.45,0) |- (act.west);
\draw[arr] (mutrepo.south) -- node[right, font=\scriptsize] {event path} (event.north);
\draw[arr] (event) -- (real);
\draw[deny] (event.east) -- ++(1.45,0) node[right, font=\scriptsize, align=left]
  {listener may suppress\\external dispatch};
\end{tikzpicture}}
\caption{Query evaluation and action dispatch in VirtualSet}
\label{fig:architecture}
\end{figure}

\subsection{Executable ontology contract}

VirtualSet uses \emph{ontology} in the operational knowledge-engineering sense of an explicit,
versioned conceptualization shared by query construction, validation, data access, and action
dispatch \cite{gruber-ontology-1993,poggi-obda-2008}. A world ontology at version $v$ is

\begin{equation}
\mathcal{O}^{v}=\langle\mathcal{C},\mathcal{F},\mathcal{E},\mathcal{K},
\mathcal{A},\mathcal{P},\lambda\rangle^{v},
\label{eq:operational-ontology}
\end{equation}

where $\mathcal{C}$ contains entity and collection concepts; $\mathcal{F}$ contains typed
attributes with domains and codomains; $\mathcal{E}$ contains typed relations with declared source,
target, and cardinality; $\mathcal{K}$ contains keys, nullability conditions, and structural
constraints; $\mathcal{A}$ contains receiver-bound operations with parameter, return, and effect
signatures; $\mathcal{P}$ identifies the policies and entitlement predicates used for grant
projection; and $\lambda$ maps concepts, attributes, and relations to repository structures. A
runtime interpretation $\mathcal{I}$ assigns extents to concepts and relations. Under
Eq.~\eqref{eq:operational-ontology}, a world is the
pair $W^{v}=(\mathcal{O}^{v},\mathcal{I})$. The action registry $R^{r}$ realizes the operation
symbols in $\mathcal{A}$ as versioned executable descriptors.

Table~\ref{tab:ontology-realization} connects this conceptual contract to the implementation and
to the QRC fields that preserve it across execution. The mapping is bidirectional at runtime:
$\lambda$ lowers ontology queries to SQL or stream evaluation, while repository rows are lifted
back into typed entity values whose stable identifiers form the query denotation. Consequently,
the authority carried by a QRC is derived from an interpretation of the same ontology that
determines the legal action receiver.

\begin{table}[!htbp]
\centering
\caption{Runtime implementation of the ontology contract}
\label{tab:ontology-realization}
\small
\setlength{\tabcolsep}{4pt}
\begin{tabular}{>{\raggedright\arraybackslash}p{1.5cm}>{\raggedright\arraybackslash}p{4.2cm}>{\raggedright\arraybackslash}p{4.7cm}}
\toprule
Ontology term & VirtualSet realization & QRC relevance \\
\midrule
$\mathcal{C}$ & Entity and collection outlines & Concrete receiver type $T$ \\
$\mathcal{F},\mathcal{E}$ & Typed fields and navigable edges & Query plan and dependency footprint $\Delta_Q$ \\
$\mathcal{K}$ & Keys, nullability, signatures, and structural constraints & Validated AST and stable entity identity \\
$\mathcal{A}$ & Receiver-bound CRUD and domain-action descriptors & Projected action, argument, and effect grants $\Omega,B$ \\
$\mathcal{P}$ & Policy and principal-entitlement descriptors & Authenticated principal and grant dependencies \\
$\lambda,\mathcal{I}$ & Repository mapping and current entity extents & Query denotation $D$ and selected-state fingerprints \\
\bottomrule
\end{tabular}
\end{table}

\subsection{Receiver closure and complete-AST validation}
\label{sec:receiver-closure}

Throughout the paper, the \emph{receiver} of a member selection or call is the value on which the
member is selected: in \texttt{employees.filter(...)}, the collection \texttt{employees} is the
receiver of \texttt{filter}, and each element \texttt{e} is the receiver of \texttt{e.age}. An
ontology world $W$ defines a set of entity and collection types
$\mathcal{T}_W$. Let $\mathcal{T}^{\mathrm{cl}}_W\subseteq\mathcal{T}_W$ denote the request-resolved
ontology receivers whose member sets are fixed by the captured world. For a receiver type
$T\in\mathcal{T}^{\mathrm{cl}}_W$, let
$\mathcal{Q}_{W^v}(T)$ denote the fields, edges, and collection operators exposed by world version
$W^v$, and let $\mathcal{A}_{W^v,R^r}(T)$ denote the receiver-bound action descriptors that are
enabled by $W^v$ and applicable in action-registry version $R^r$. A request contract
$S=(W^v,R^r)$ identifies both versions and determines the complete legal receiver closure

\begin{equation}
\mathcal{M}_S(T) \triangleq
\mathcal{Q}_{W^v}(T) \cup \mathcal{A}_{W^v,R^r}(T).
\label{eq:receiver-closure}
\end{equation}

The closure in Eq.~\eqref{eq:receiver-closure} binds ontology and registry versions to the
admissible member set of each closed receiver.

Validation operates on the complete expression AST, including field selection, invocation,
lambda, tuple, conditional, cast, and block nodes;
$\Gamma\vdash_S e:T$ denotes recursive validation of expression $e$ in lexical environment
$\Gamma$ under that contract. If a subexpression $e$ has a closed ontology receiver type $T$, selection of a member
$m$ is structurally valid only if $m$ belongs to the closure of that receiver:

\begin{equation}
\frac{\Gamma\vdash_S e:T \qquad T\in\mathcal{T}^{\mathrm{cl}}_W
      \qquad m:\tau\in\mathcal{M}_S(T)}
     {\Gamma\vdash_S e.m:\tau[\,T/\mathtt{this}\,]} .
\label{eq:receiver-selection}
\end{equation}

The substitution in Eq.~\eqref{eq:receiver-selection} preserves type information across fluent
collection operators. When an inherited operator declares a future-\texttt{this} result, its
return type resolves to the concrete collection receiver rather than only to the generic base
type. The subsequent selection is therefore validated against the same collection-specific
closure. An action terminal follows the same selection rule and additionally requires arguments
compatible with its receiver-bound signature before dispatch.

Validation recursively checks every child. A lambda body uses $\Gamma[x\mapsto A]$, with $A$
projected from its higher-order operator; captured variables resolve through $\Gamma$.
World/module compilation validates user-defined bodies and installs their signatures atomically
in $W^v$. Per-request validation reads this environment without modifying $W^v$ or $R^r$.
GCP supplies the general unification rules \cite{dli-paper}; the world-indexed closure is the
structural premise used here for authority compilation.

\textbf{Proposition 1 (AST-wide closed-receiver exclusion).} Let
$S_{\mathrm{val}}=(W^v,R^r)$ be the stable
issue-time contract captured for AST validation and dependency projection. If recursive validation accepts a VirtualSet
expression AST $e$, then for every member-selection node $e_0.m$ whose inferred receiver satisfies
$\Gamma\vdash_{S_{\mathrm{val}}} e_0:T_0$ and $T_0\in\mathcal{T}^{\mathrm{cl}}_W$, $m$ belongs to
$\mathcal{M}_{S_{\mathrm{val}}}(T_0)$; every invocation of a selected receiver member has arguments validated
against the parameter types of its projected signature. When a controller issues a QRC,
it projects the requested actions separately, after validation of the selecting query.

\textit{Proof sketch.} Structural induction over $e$ establishes the result. For a closed-receiver
selection, the hypothesis validates $e_0$ and Eq.~\eqref{eq:receiver-selection} establishes
membership and result substitution. Invocation projects the callee signature before checking
arguments sequentially. A lambda expected at $A\mathbin{\to}B$ is checked under
$\Gamma[x\mapsto A]$ with result $B$. Tuple, conditional, cast, and block cases apply the
hypotheses to all children; user-defined calls use signatures accepted in $W^v$.
Other values use Outline's ordinary member rules. This establishes query legality in
$S_{\mathrm{val}}$; the later-view authority condition is established separately by Theorem~1.

\subsection{Compositional receiver typing}
\label{sec:type-continuity}

For a collection receiver $C\langle A\rangle$, $A$ is the element type and $C$ is the concrete
interface, including domain extensions. GCP projects each called signature against its receiver
and arguments \cite{dli-paper}. A future-\texttt{this} result preserves the concrete interface;
a transforming result establishes the interface appropriate to its new type.
Table~\ref{tab:receiver-transitions} gives the ontology transitions used by the dependency
analysis. Supplementary Material, Sections~S3 and~S7, provides the underlying language examples.

\begin{table}[!htbp]
\centering
\caption{Receiver types across collection operations}
\label{tab:receiver-transitions}
\small
\setlength{\tabcolsep}{3pt}
\begin{tabular}{>{\raggedright\arraybackslash}p{2.6cm}>{\raggedright\arraybackslash}p{3.5cm}>{\raggedright\arraybackslash}p{4.8cm}}
\toprule
Operation class & Representative transition & Information passed onward \\
\midrule
Receiver preserving & $C\langle A\rangle\to C\langle A\rangle$ & \texttt{filter}, \texttt{order\_by}, and \texttt{take} retain the concrete collection and element type. \\
Domain navigation & $C\langle A\rangle\to D\langle B\rangle$ or $A\to B$ & The declared edge target establishes a new receiver and its domain members. \\
Element projection & $C\langle A\rangle\to V\langle B\rangle$ & \texttt{map} checks $A\to B$ and propagates $B$; source collection extensions are not automatically transferred. \\
Grouping & $C\langle A\rangle\to\mathrm{GroupBy}\langle K,A\rangle$ & Key type $K$ and element type $A$ remain distinct inputs to grouped operations. \\
Generic collection result & $C\langle A\rangle\to V\langle A\rangle$ & \texttt{distinct} and set operations retain element information under their declared generic result interface. \\
Materialization & $C\langle A\rangle\to A$, $[A]$, or $\mathrm{Int}$ & \texttt{first}, \texttt{to\_list}, and \texttt{count} establish entity, list, or scalar receivers. \\
\bottomrule
\end{tabular}
\end{table}

Let $\Pi_S(\tau,m,u)$ be the result type obtained by checking $u$ against member $m$ projected
from receiver type $\tau$ under request contract $S$. A callback parameter is bound to the
projected element type before its body is inferred. The body result instantiates the result
parameter of a transforming operator. For a chain prefix $e_i=e_{i-1}.m_i(u_i)$, define

\begin{equation}
\tau_0=\operatorname{type}_S(e_0),\qquad
\tau_i=\Pi_S(\tau_{i-1},m_i,u_i),\qquad
\operatorname{next}_S(e_i)=\mathcal{M}_S(\tau_i)
\label{eq:type-continuity}
\end{equation}

whenever $\tau_i$ is a closed ontology receiver. For scalar, list, and other structural results,
\texttt{next} uses the corresponding ordinary Outline member rule. A local binding records
$\tau_i$ in $\Gamma$; subsequent use reads that type rather than the type of the original root.
An explicit widening or a branch join follows its own declared typing rule. The property studied
here is the absence of additional receiver erasure at operator boundaries that declare
preservation, together with faithful propagation of declared result changes.

\textbf{Proposition 2 (Compositional receiver continuity).}
Fix a request contract $S$ and accepted operator declarations. For a well-typed composition whose
calls use the contracts in Table~\ref{tab:receiver-transitions}, every prefix has the result type
given by Eq.~\eqref{eq:type-continuity}. Along a segment of receiver-preserving calls, the concrete
collection interface and its applicable member signatures remain available. A transforming or
materializing call establishes a new result type, and subsequent lookup uses that result's
interface. Local bindings and callbacks transmit their projected types under the same rules.

\textit{Proof.} Induct on the checked composition. Roots and variables obtain their types from
$S$ and $\Gamma$. Equation~\eqref{eq:receiver-selection} substitutes the concrete receiver for
preserving results; signature instantiation supplies the declared type for each other transition.
Bindings and callbacks transmit these types through $\Gamma$, so each subsequent lookup uses
the preceding expression's result.

Prefix types determine dependency owners in $\Delta_Q$; the evaluated selecting result supplies
the entity type and identifiers used by QRC. This connects structural applicability to the
result-specific authority compiled in Section~\ref{sec:model}.

\subsection{Running query-to-action workflow}
\label{sec:running}

The running example uses the same employee selection for an analytical report and a
subsequent recovery request. The initial request asks:

\begin{quote}
How many retirement-eligible employees are still holding assigned company laptops whose return
date is missing, and what is their total recorded laptop value?
\end{quote}

In this workflow, the supplied business rule defines retirement eligibility as an age
of at least 65 years. A generated expression might reference \texttt{returned = false} even though
the schema defines \texttt{returnDate}, or it might apply an exclusive age boundary. The first is
a structural error that the receiver guard can reject before execution; the second is a semantic
error whose resolution depends on domain evidence or metadata.

Listing~\ref{lst:risk-query} names the selected employee collection \texttt{at\_risk}.
The count and total are the values shown to the requester; neither is an action receiver.

\begin{lstlisting}[float=htbp,caption={Receiver-typed analytical selection.},label={lst:risk-query}]
let at_risk = employees
  .filter(e -> e.age >= 65)
  .filter(e -> e.assignedLaptop().id != null
            && e.assignedLaptop().returnDate == null);

(at_risk.count(),
 at_risk.sum(e -> e.assignedLaptop().value))
\end{lstlisting}

Receiver validation rejects undeclared fields and navigation methods. The explicit guard
on the navigated laptop identifier excludes both null
and dangling links: an absent edge and its field reads evaluate to null, so testing only
\texttt{returnDate == null} would include employees without a laptop. SQL lowering preserves
this rule through correlated scalar subqueries.

A subsequent request concerns the employees rather than the two aggregates:

\begin{quote}
Initiate laptop recovery for those employees.
\end{quote}

Let $Q$ denote the filter expression defining \texttt{at\_risk}. To approve recovery,
the controller submits $Q$ and the requested action to the issuer. If policy grants the
action, the issuer binds it to $D$, the employee identifiers selected when approval is
issued. Listing~\ref{lst:risk-action} separates issuance from the later invocation in
the notation of Algorithm~\ref{alg:qrc-dispatch}.

\begin{lstlisting}[float=htbp,caption={Approval and delayed recovery for selected employees},label={lst:risk-action}]
requested = {initiate_recovery: {}}
C = IssueQRC(Q, p, requested, lifetime, nonce)

I = (p, Employee, C.q, A, initiate_recovery, {}, effect)
Dispatch(C, I)
\end{lstlisting}

The caller presents the original capability $C$ and requests a nonempty subset
$A\subseteq D$ with an effect permitted by the grant. Even if another employee now meets
the filter, the caller cannot add that employee to $A$. A policy may recheck current
eligibility, but that check cannot enlarge the original approval. Local mutation and
subsequent external delivery have distinct transaction boundaries (Section~\ref{sec:events}).

The continuation policy determines whether approval is historical or conditional on current
eligibility. The default profile binds selected employee rows and referenced ontology declarations.
For recovery that must stop when a laptop is returned, the functional-navigation profile also
binds the link to each selected employee's laptop and the return-date field read there.
A return then invalidates the continuation,
while changes outside the witness closure remain admissible. Section~\ref{sec:scope} defines
these G1 and G2 contracts, and Theorem~2 establishes the stronger profile for its query fragment.

\subsection{Problem statement and design requirements}
\label{sec:problem}

Given an accepted ontology query $Q$ over contract $S=(W^v,R^r)$ with receiver $T$,
denotation $D$, and an approval for principal $p$, the \emph{authority-continuity
problem} is to ensure that any later
state-changing invocation that claims $Q$ as its justification (i) addresses only objects in $D$,
(ii) invokes only actions and effects granted to $p$ on $T$ at issuance and
whose recorded authorization dependencies remain valid at dispatch,
(iii) executes under the same authenticated principal, and (iv) enters the batch with
matching issue-time fingerprints for the recorded ontology, policy, and executor
dependencies, the invoked target rows, and any profile-specific query witnesses.
Finally, (v) revisions outside this active footprint leave the dependency check unchanged.

\textbf{Definition (receiver-indexed authority compilation).} Let an accepted query $Q$ under
contract $S$ evaluate to receiver $T$, denotation $D$, and receiver-qualified fingerprint map
$\Delta_Q$. Let $\mathcal{R}_{\mathrm{req}}$ map the requested actions to their approved
arguments; projecting this request for principal $p$ yields grants $\Omega$ and argument
bounds $B$. For witness profile $\chi$, define
$\Delta=\Delta_Q\cup\operatorname{StateDeps}_S(Q,T,D,\chi)
\cup\operatorname{Deps}_S(\Omega,p)$. Receiver-indexed authority compilation is the judgment
\begin{equation}
\operatorname{Compile}_{S,p}(Q,\mathcal{R}_{\mathrm{req}},\chi)\Downarrow
  C\langle T,D,\Omega,B,\Delta\rangle ,
\label{eq:authority-compilation}
\end{equation}
whose dispatch interpretation admits an invocation only through the same receiver $T$, within
$D$, under $\Omega$ and $B$, while the invocation's active recorded dependencies remain valid.
Equation~\eqref{eq:authority-compilation}
is the semantic bridge between compositional typing and QRC: the result of query validation and
evaluation becomes the authority object consumed by the action runtime.

This formulation imposes five requirements on the model-facing interface. First, the world loaded
for the current request serves as the type authority, thereby accommodating tenant-specific
schemas. Second, each intermediate receiver determines the legal member set used for validation.
Third, operators declared to return future-\texttt{this} preserve collection-specific
extensions; other operators propagate their declared result receiver. Fourth, the
representation supports both set-oriented reads and explicitly effectful terminals, with
receiver validation completed before repository mutation. Fifth, query execution yields the
denotation and dependency footprint needed to attenuate action authority; dispatch revalidates
those exact dependencies rather than relying on an unscoped identifier supplied by the caller.
Section~\ref{sec:model} develops the capability that meets requirements (i)--(v);
Section~\ref{sec:scope} defines the runtime state bound by each continuity profile.

\section{Query-Derived Receiver Capability Model}
\label{sec:model}

This section defines the query-derived receiver capability. Section~\ref{sec:authority} derives
the authority envelope from the evaluation judgment of an accepted query;
Section~\ref{sec:trust} states the trust model, the two integration modes in which a capability
is exercised, and its lifecycle; Section~\ref{sec:dispatch} gives the issuance and dispatch
algorithm and proves the authority-preservation theorem; Section~\ref{sec:scope} specifies the
continuity profiles and their action-entry enforcement.

\subsection{Query-derived receiver authority}
\label{sec:authority}

Throughout, $T$ denotes an \emph{entity} type of the world and $\mathrm{VirtualSet}\langle
T\rangle$ its typed collection. An \emph{issuable} query is an accepted program with exactly one
collection root whose result is still a collection of that root's entity type---a chain of
root-entity-preserving operators (filtering, ordering, bounding, distinctness) whose predicates may
navigate edges and read fields of other types but may not call actions, effectful operators, or
free functions, and whose chain the relational translator lowers completely;
Section~\ref{sec:execution} states the admissibility check. For an issuable query $Q$ with receiver
$T$, VirtualSet records the evaluation judgment
\begin{equation}
\Gamma\vdash_S Q:\mathrm{VirtualSet}\langle T\rangle
\Downarrow (D,\Delta_Q),
\label{eq:query-evaluation}
\end{equation}
where $D$ is the finite set of returned identifiers of objects of type $T$ and $\Delta_Q$ maps
every field and edge referenced during query validation and evaluation, under the receiver that
owns it, to a fingerprint of its recorded schema attributes (Table~\ref{tab:dependency-map}).
Write $K_Q$ for the schema keys derived from the validated query. At issuance,
$\operatorname{dom}(\Delta_Q)=K_Q$ and $\Delta_Q(d)=S(d)$ for each $d\in K_Q$.
QRC authority is stated at entity granularity: a
capability names a set of $T$ objects, and a dispatch applies an action registered on receiver
$T$ to each object in its target set. The judgment uses the source operator semantics;
an execution backend must preserve this denotation. The cross-backend expression tests in
Section~\ref{sec:equal-g2} check this obligation from reference expressions through selected
identifiers to repository mutations. Let
$\mathcal{R}_{\mathrm{req}}$ map each requested action name to its controller-approved argument map.
The trusted controller projects the actions, parameter types, and effects authorized for principal
$p$ on receiver $T$ under $S$:
\begin{equation}
\begin{aligned}
\Omega&=\operatorname{ProjectGrants}_S
  (p,T,h(Q),D,\operatorname{dom}(\mathcal{R}_{\mathrm{req}})),\\
\overline{x}_a&=\mathcal{R}_{\mathrm{req}}(a),\quad a\in\operatorname{dom}(\Omega),\\
B(a)&=h_{\mathrm{arg}}(\overline{x}_a),\\
\Delta&=\Delta_Q\cup\operatorname{StateDeps}_S(Q,T,D,\chi)
       \cup\operatorname{Deps}_S(\Omega,p).
\end{aligned}
\label{eq:qrc-projection}
\end{equation}
Argument maps are normalized into immutable value trees at issuance and invocation entry,
before either operation waits for runtime locks. The maps in Eq.~\eqref{eq:qrc-projection} and
the invocation below denote these frozen values. Each projected signature $\sigma$ records
parameter types, required versus optional presence, and its descriptor-derived calling
convention. Optional presence permits omission; null is admitted only by a nullable type.
The calling convention distinguishes a positional value, a named-parameter container, and an
Outline record independently of the number of supplied fields. These signature components are
authenticated with $\Omega$, so neither omission rules nor argument interpretation can be
changed by a continuation caller.
The unions in Eq.~\eqref{eq:qrc-projection} combine fingerprint maps captured in the same
view $S$, so any shared key has the same value in both maps. The state map covers full-row
keys for $D$ and, for either witness profile, the field-state keys derived from $Q$.
The grant map covers the applicable action, policy, principal-entitlement and executor keys.
Issuance requires $D\ne\emptyset$,
$\Omega\ne\emptyset$, and a positive lifetime; an empty result carries no object-level authority.
The issuer then
constructs the canonical payload and authentication tag
\begin{equation}
\begin{aligned}
P&=\langle c,\iota,w,p,T,h(Q),D,\Omega,B,\Delta,\rho,\mathcal{J},n,\chi,Q_r\rangle,\\
\mu&=\operatorname{HMAC}_{k_\iota}(\operatorname{enc}(P)),\qquad C=\langle P,\mu\rangle,
\end{aligned}
\label{eq:qrc}
\end{equation}
Equation~\eqref{eq:qrc} defines the authenticated QRC envelope used by the dispatcher, where
$c$ is the capability identifier, $\iota$ the issuer identifier, and $w$ the identifier of the
\emph{world}---the ontology, repository, and action registry---from whose state $S$ the
capability is minted, $\rho=\langle r_W,r_R\rangle$ the issue-time world and action-registry
audit revisions, and $\operatorname{enc}$ an injective, length-delimited canonical encoding.
The implementation authenticates this encoding with HMAC-SHA-256. An
issuer $(\iota,k_\iota)$ is bound to exactly one world: it mints only for $w$ and verifies only
in $w$. The half-open validity interval $\mathcal{J}=[t_{\mathrm{iss}},t_{\mathrm{exp}})$,
where $t_{\mathrm{exp}}>t_{\mathrm{iss}}$, bounds the instant at which the capability may be
\emph{accepted}, and the issuer--nonce pair $(\iota,n)$ supports atomic at-most-once
consumption. We call $P$ the
\emph{authority envelope}: its denotation and grants bound what a continuation may address and do,
while its provenance and dependency fields bind the context that established that authority.
The authenticated profile $\chi$ identifies selected rows, functional closure, finite typed paths,
or protected query rechecking. $Q_r$ is the validated source of $Q$ for rechecking and is empty otherwise; when present,
its canonical digest must equal the recorded plan digest, denoted $C.q=h(Q)$.
The authenticated profile attests to successful issue-time classification:
$g(\chi)=G1$ for selected rows and $g(\chi)=G2$ for either witness profile or protected rechecking.
Write $\mathbf{s}_\chi=(g(\chi),e(\chi))$ for a trusted profile's strength and
$\mathbf{m}_{S}(T,a)=(g_{\min}^{S}(T,a),e_{\min}^{S}(T,a))$ for the action requirement.
The ordering is componentwise: the first coordinate orders continuity levels; the second
minimum is 1 when equality of the approved query's observations is required, and 0 otherwise.
A profile earns $e(\chi)=1$ only if it captures and revalidates at least those observations.
A Java descriptor defaults to $e_{\min}=0$ and can require $e_{\min}=1$ independently of its
continuity minimum. Its fingerprint binds both settings. The independent SQL comparison
implements the same two policy coordinates in SQLite.
Global revisions are retained for audit, whereas dispatch validity depends on the members in
$\Delta$. This separation assigns grant construction to the application controller and
cryptographically limits subsequent callers to the controller-authorized projection. Because
$\operatorname{ProjectGrants}$ receives $h(Q)$ and $D$, policy may attenuate actions by both query
provenance and selected objects rather than by receiver type alone.

An invocation request is
$I=\langle p',T',q,A,a,\overline{x},\omega\rangle$, where $q$ is the submitted canonical-plan
digest, $A$ the target set, $a$ the action, $\overline{x}$ the arguments, and $\omega$ the requested
effect. The dispatcher obtains $p'$ from the
authenticated execution context rather than from model-produced arguments. For an invocation
target set $A$, let $\Delta\!\downarrow_A$ retain every recorded dependency except a full-row
$\textsc{EntityState}(T',id)$ key whose $id\notin A$:
\[
\operatorname{dom}(\Delta\!\downarrow_A)=
\{d\in\operatorname{dom}(\Delta)\mid
\neg\exists id\;[id\notin A\land d=\textsc{EntityState}(T,id)]\},
\]
with retained keys inheriting their values from $\Delta$. Issuance records a full-row key for every member
of $D$, allowing any later subset to be selected, while dispatch makes the full-row freshness
premise specific to the objects that will actually receive the action. Receiver-qualified
field-state witnesses remain active because they encode the data dependence of the accepted query.
This restriction applies to the root-row keys introduced for $D$; the evaluated automatic
profiles introduce no full-row keys for other receiver types. The field-state closure is retained
for all selected roots. Full-row freshness is therefore target-specific, while functional
witness validation covers the shared issue-time closure. A change to a field-state witness
used by a non-invoked root can consequently reject the continuation.
Authorization succeeds
at the logical verification boundary exactly when
\begin{align}
\operatorname{Auth}(C,I,S') \Longleftrightarrow{}&
 \operatorname{TrustedIssuer}(\iota) \land \operatorname{VerifyMAC}_{k_\iota}(C)
 \land w=w'
 \nonumber\\
&{}\land |\operatorname{dom}(\Delta)|\le\beta
 \land \operatorname{fresh}(\iota,n) \land t_{\mathrm{acc}}\in\mathcal{J}
 \nonumber\\
&{}\land p'=p \land T'=T \land q=h(Q) \land \emptyset\ne A\subseteq D
 \nonumber\\
&{}\land a:\sigma/\mathcal{E}\in\Omega
 \land \overline{x}:\sigma \land h_{\mathrm{arg}}(\overline{x})=B(a)
 \land \omega\in\mathcal{E}
 \nonumber\\
&{}\land S_0\equiv_{\Delta\downarrow_A}S'
 \land \mathbf{s}_{\chi}\succeq\mathbf{m}_{S'}(T,a)
 \nonumber\\
&{}\land \operatorname{Evidence}_{\chi}(C,A,S'),
\label{eq:qrc-auth}
\end{align}
where $\beta$ is the verifier's bound on the number of recorded dependency keys (a
configuration constant shared by issuer and verifier; the issuer refuses to mint a capability
whose $\Delta$ exceeds it, so every issued capability satisfies this premise and it can fail only
for a payload the issuer did not produce), $w'$ is the world in which the verifier runs,
$t_{\mathrm{acc}}$ is the \emph{acceptance instant}---the instant at which $(\iota,n)$ enters the ledger, read under the
ledger's exclusion---and $S_0\equiv_{\Delta\downarrow_A}S'$ means that every active key has the
same recorded fingerprint in the issue-time and dispatch-time views. This relation includes each
target entity's issue-time row state, all receiver-qualified query witnesses, and the recorded
schema, action, policy, principal, and executor dependencies. The world
premise is what prevents a capability minted in one world from being accepted by another world
that shares the issuer configuration, receiver types, identifiers, and fingerprints; the
acceptance-instant premise fixes \emph{which} instant the lifetime constrains, since verification,
locking, and the durable ledger append take time. After authorization, the dispatcher constructs the runtime value
directly from the same $\overline{x}$ whose type and digest were checked; no second payload input is
accepted by the execution API. The minimum $\mathbf{m}_{S'}(T,a)$ is read from the live descriptor;
meeting a minimum selects an eligible profile, whereas the evidence premise tests the actual state.
For selected-row, closure, and typed-path profiles, $\operatorname{Evidence}_{\chi}(C,A,S')$
uses the trusted issuer's authenticated classification and the already checked footprint;
it does not require the original query at dispatch. For recheck it requires
$h(C.Q_r)=C.q$, revalidates $C.Q_r$ as an admissible selection of $C.T$, and checks
$A\subseteq[\![C.Q_r]\!]_{S'}$ under a repository view stabilized through action entry.
Section~\ref{sec:scope} specifies the profile requirements.
Equation~\eqref{eq:qrc-auth} specifies logical authorization. Operational dispatch additionally
requires successful repository capture and transaction establishment; failures at either step
stop execution before action entry, including failures while capturing inactive keys.
The subset premise is the essential query-to-action link: possession of a valid receiver-level
action name cannot enlarge the query result. The dependency premise separates relevant from
unrelated ontology evolution. A change to an unreferenced member may advance the global world
revision while leaving every active fingerprint unchanged, preserving the dependency check.

\subsection{Trust model, integration modes, and capability lifecycle}
\label{sec:trust}

In \emph{coupled} integration, the controller issues the capability and submits it for
dispatch in a single call. In \emph{decoupled} integration, the client retains the
capability and presents it later. In both modes, issuance and dispatch use separate
repository transactions; the dispatcher revalidates the capability before action entry.
The adversary controls
query text and the requested continuation; in the latter mode it may also alter, substitute,
or replay the serialized capability. The covered substitutions concern receiver, targets,
query provenance, principal, operation, arguments, effects, dependencies, and lifetime.
The evaluation exercises both integration modes.

The trusted base comprises validation and evaluation, grant projection, authenticated-principal
resolution, the issuer key, dependency-version providers, the nonce ledger, and the stable
dispatcher/repository path. Application policy attenuates the projected grants. The execution
principal comes from the authenticated context and must match the invocation principal.
The formal properties assume HMAC unforgeability, collision-resistant canonical digests,
issuer-unique nonces, a monotonic trusted clock, and request-stable dependency versions.

The link to \emph{current} policy rests on a contract with the grant provider. Fix the
principal, receiver type, query digest, approved identifiers, and requested actions in
Eq.~\eqref{eq:qrc-projection}. If the policy would cease to authorize an issued grant
with its recorded signature and effects, at least one recorded authorization
fingerprint must differ from its issue-time value. This condition applies also to external
inputs used by policy: for example, a spending limit that determines a grant must be
reflected in the policy or entitlement fingerprint. Under this contract, validating the
recorded dependencies preserves the original grant under current policy. Dispatch checks
those fingerprints and the live descriptor; grant projection itself takes place at issuance.

Interpreting a preserved observation also requires stable read semantics. For each field
or link read by the query, a deployment must either preserve its meaning for outstanding
approvals or change a compared fingerprint when that meaning changes. For example, a new
interpretation of a scalar with the same SQL representation requires a new executor version
if the field key itself is unchanged. The verifier compares the versions supplied by the
deployment; it does not infer a change of meaning from the stored value.

During dispatch, the world monitor, the registry-wide read lock, and a repository write
transaction protect the interval from dependency capture through local action completion. PostgreSQL locks every
recorded state row with \texttt{SELECT FOR UPDATE}; SQLite uses \texttt{BEGIN IMMEDIATE}.
Policy and executor providers supply request-stable versions. The world and registry locks
coordinate one process per world; multi-process deployment additionally coordinates those
configuration views.

The ledger admits each issuer--nonce pair once. Acceptance time is read under ledger
exclusion; an expired request is rejected there, while an accepted batch may finish
after its deadline. Fingerprints compare the issue and dispatch views: restoring a
recorded value restores that premise. Irreversible revocation uses a monotone version
or nonce revocation. The file ledger and the business database commit separately;
Supplementary Sections~S5 and S13 give the storage and recovery protocols.

If dispatch fails before the ledger consumes the nonce, the same capability may be
presented again and is revalidated against the current state. This includes a PostgreSQL
serialization failure while acquiring a row lock. Once the ledger consumes the nonce,
targets execute in ascending identifier order with one savepoint per target. A recoverable
action or synchronous-subscriber failure rolls back that target and commits the completed
prefix; a transaction-level failure rolls back the whole batch. Neither rollback restores
the spent nonce. Retrying after consumption requires a new capability issued from the
current query result. A completed-target list or durable business idempotency key governs
repetition across capabilities. Workflows requiring all-or-nothing treatment of
recoverable target failures add a batch-transaction policy.

\subsection{Issuance, stable dispatch, and authority preservation}
\label{sec:dispatch}

Algorithm~\ref{alg:qrc-dispatch} gives the Java protocol in the stable regions of
Section~\ref{sec:trust}, with its action checks written in the componentwise notation of
Eq.~\eqref{eq:qrc-auth}. The Java permit carries both the verified continuity level and the issuer profile's evidence
strength. The independent SQLite prototype implements the same policy check with its own
transactional ledger; the locks, ledger, and savepoints shown here are those of Java. Before acquiring locks or reading the repository, dispatch verifies
the capability's tag, issuer, world, and dependency-key bound $\beta$. Issuance enforces the
same bound on line~9. The \textsc{Admissible} check establishes a read-only, root-entity-preserving
selection and the configured profile's fragment or recheck-adapter support. The trusted service
fixes $\chi$; \textsc{SourceForProfile} retains the validated query source for rechecking.

Query evaluation and fingerprint capture observe one issue-time snapshot: PostgreSQL uses
\texttt{REPEATABLE READ}, and SQLite uses a serialized connection with \texttt{BEGIN IMMEDIATE}.
Dispatch captures recorded rows under locks. A PostgreSQL serialization failure while acquiring
a row lock aborts capture before nonce consumption; a fresh presentation then observes the
committed state. The algorithm groups authorization premises logically; the implementation checks
the live effect and minimum before capture. The dispatcher compares the active dependency
fingerprints inside its transaction before the ledger consumes the nonce. At line~25,
the frozen input tree and authenticated signature form an immutable argument
plan. Decoding is checked before consumption, and \textsc{ConsumeFor} materializes a fresh value
for each target, preserving the approved input across caller and earlier-target mutations.

Capture and comparison have different scopes. The evaluated dispatcher captures all keys in
$\Delta$ before comparing $\Delta\!\downarrow_A$; it does not implement target-pruned I/O or
locking. A missing entity row is represented by a distinguished fingerprint, not a capture
exception. Its changed full-row fingerprint therefore rejects a targeted object, but does not
reject a non-targeted root solely because that root disappeared. Field-state witnesses remain
active. This behavior motivates the target-scoped statement of Theorem~2 and explains why
dispatch cost can grow with $|D|$ even when $|A|=1$.

\begin{algorithm}[!htbp]
\caption{QRC issuance and stable dispatch.}
\label{alg:qrc-dispatch}
\fontsize{10}{10.6}\selectfont
\begin{algorithmic}[1]
\Procedure{IssueQRC}{$Q,p,\mathcal{R}_{\mathrm{req}},\ell,n$}
  \State acquire world monitor; begin repository write transaction;
    acquire registry-wide action read lock
  \State $\chi\gets\mathit{issuer.witnessMode}$; $S\gets\Call{CaptureRuntimeView}{}$;
    require $\Call{Validate}{Q,\Gamma_S,S}$ and $\Call{Admissible}{Q,S,\chi}$
  \State $(T,K_Q)\gets\Call{ReceiverAndKeys}{Q,S}$;
    $D\gets\Call{EvaluateIds}{Q,p,T,S}$; require $D\ne\emptyset$
  \State $\Omega\gets\Call{ProjectGrants}{S,p,T,h(Q),D,\operatorname{dom}(\mathcal{R}_{\mathrm{req}})}$
    and require $\Omega\ne\emptyset$, $\ell>0$, and
    $\mathbf{s}_\chi\succeq\mathbf{m}_{S}(T,a)$ for each $a\in\operatorname{dom}(\Omega)$
  \State require every $\mathcal{R}_{\mathrm{req}}(a)$ conforms to $\Omega[a].\sigma$
    for $a\in\operatorname{dom}(\Omega)$
  \State $B\gets\{a\mapsto h_{\mathrm{arg}}(\mathcal{R}_{\mathrm{req}}(a))
    \mid a\in\operatorname{dom}(\Omega)\}$
  \State $Q_r\gets\Call{SourceForProfile}{Q,\chi}$;
    $K\gets K_Q\cup\Call{StateKeys}{S,Q,T,D,\chi}$
  \State $K\gets K\cup\Call{GrantKeys}{S,\Omega,p}$;
    $\Delta\gets S|_K$; require $|K|\le\beta$
  \State $\rho\gets\Call{AuditRevisions}{S}$; $c\gets\Call{CapabilityId}{\iota,n,h(Q)}$;
    $t_{\mathrm{iss}}\gets\Call{Now}{}$; $\mathcal{J}\gets[t_{\mathrm{iss}},t_{\mathrm{iss}}+\ell)$
  \State $P\gets\langle c,\iota,w,p,T,h(Q),D,\Omega,B,\Delta,\rho,\mathcal{J},n,\chi,Q_r\rangle$;
    $\mu\gets\operatorname{HMAC}_{k_\iota}(\operatorname{enc}(P))$
    \Comment{$w$: identifier of this world}
  \State commit repository transaction; release locks
  \State \Return $C=\langle P,\mu\rangle$
\EndProcedure
\Procedure{Dispatch}{$C,I$}
  \State $p'\gets\Call{AuthenticatedPrincipal}{}$; require $p'=I.p$
  \State require $|\operatorname{dom}(C.\Delta)|\le\beta$, $\Call{VerifyMAC}{C}$,
    $\Call{TrustedIssuer}{C.\iota}$, $C.w=w$ \Comment{envelope}
  \State acquire world monitor; begin repository write transaction;
    acquire registry-wide action read lock
  \State $S'\gets\Call{CaptureAndLockDependencies}{\operatorname{dom}(C.\Delta)}$
  \State require $\Call{Fresh}{C.\iota,C.n}$
  \State require $I.p=C.p$, $I.T=C.T$, and $I.q=C.q$
  \State require $\emptyset\ne I.A\subseteq C.D$
  \State require $I.a\in\operatorname{dom}(C.\Omega)$,
    $I.\overline{x}:C.\Omega[I.a].\sigma$, and $I.\omega\in C.\Omega[I.a].\mathcal{E}$
    and $h_{\mathrm{arg}}(I.\overline{x})=C.B(I.a)$
  \State require $\forall d\in\operatorname{dom}(C.\Delta\!\downarrow_{I.A}),\;S'(d)=C.\Delta(d)$,
    $\mathbf{s}_{C.\chi}\succeq\mathbf{m}_{S'}(I.T,I.a)$, and $\operatorname{Evidence}_{C.\chi}(C,I.A,S')$
  \State $v\gets\langle I.\overline{x},C.\Omega[I.a].\sigma\rangle$;
    require $\Call{DecodeCheckedArguments}{v}$ succeeds
  \State under the ledger's exclusion: $t_{\mathrm{acc}}\gets\Call{Now}{}$;
    require $t_{\mathrm{acc}}\in C.\mathcal{J}$; consume $(C.\iota,C.n)$
  \State $\pi\gets\Call{VerifiedPermit}{I.p,I.T,I.A,I.a,v,\mathbf{s}_{C.\chi}}$
  \ForAll{$d\in I.A$ in ascending identifier order}
    \State $s\gets\Call{Savepoint}{}$; $v_d\gets\Call{ConsumeFor}{\pi,p',I.T,d,I.a,\mathbf{m}_{S'}(I.T,I.a)}$;
      $r_d\gets\Call{LoadReceiver}{I.T,d}$
    \State dispatch $I.a$ on receiver $r_d$ with argument $v_d$; \Call{Release}{s}
    \State \textbf{on recoverable failure of line~30:} \Call{RollbackTo}{s}; record $d$ as failed;
      \textbf{exit loop}
  \EndFor
  \State commit repository transaction; release locks;
    \Return the set of completed targets and the failed target, if any
  \State \textbf{on transaction failure anywhere} (incl.\ savepoint failure): roll back the
    repository transaction; release locks; propagate
\EndProcedure
\end{algorithmic}
\end{algorithm}

\textbf{Proposition 3 (receiver and target confinement).}
If Algorithm~\ref{alg:qrc-dispatch} authorizes invocation $I$ under capability $C$ issued from
Eq.~\eqref{eq:query-evaluation}, every dispatched object has runtime type $T$ and identifier in
$D$.

The dispatcher constructs receivers only from checked pairs $(T,id)$ with $id\in A\subseteq D$.

\textbf{Proposition 4 (dependency-safe evolution).}
Let $S_0$ be the issue-time snapshot and $S_1$ the dispatch-time snapshot. For invocation target
set $A$, a QRC remains dependency-valid across $S_0\rightarrow S_1$ exactly when
$S_0\equiv_{\Delta\downarrow_A} S_1$,
provided that the remaining authorization premises in Eq.~\eqref{eq:qrc-auth},
including the continuity requirement and profile evidence, remain satisfied. Changes outside
$\operatorname{dom}(\Delta\!\downarrow_A)$ do not alter the dependency-validity decision.
Moreover, an accepted dispatch enters action code for its first
target under a view that still agrees with $S_0$ on $\operatorname{dom}(\Delta\!\downarrow_A)$, and no writer
other than the dispatched action invocations themselves changes a key in
$\operatorname{dom}(\Delta\!\downarrow_A)$ before commit.

The active-key comparison establishes equality. The registry lock, repository transaction,
world monitor, and stable providers exclude competing changes through commit; the batch
observes its own writes.

This criterion concerns the \emph{original} capability. If a relation is retargeted,
an approval that recorded that edge loses its basis even when the new endpoint has
the same key and value. An approval that stopped before the edge can remain valid:
its referenced declarations, observations and grant are unchanged. A new approval
issued against the successor ontology answers a different question. We separate
the two times of issuance in Tables~\ref{tab:owner-intervention} and~\ref{tab:old-approval}.

\textbf{Proposition 5 (authenticated at-most-once use).}
Assuming HMAC unforgeability, issuer-unique nonces, and an atomic nonce ledger, alteration of any
authority-bearing QRC field is rejected, at most one dispatch can consume a given issuer--nonce
pair and begin action execution, and it does so at an acceptance instant inside $\mathcal{J}$.

The tag covers the payload. The ledger checks the time and inserts $(\iota,n)$ under the
same lock. Section~S10 of the supplement proves Propositions~3--5.

\textbf{Theorem 1 (query-to-dispatch authority preservation).}
Let $C$ be issued from an accepted query judgment in Eq.~\eqref{eq:query-evaluation} and a trusted
grant projection in Eq.~\eqref{eq:qrc-projection}. Under the trust assumptions above, every
execution of \textsc{Dispatch} in Algorithm~\ref{alg:qrc-dispatch} that enters action code
(i) retains the authenticated issuer, world, principal, and query provenance; (ii) selects a
nonempty target set contained in $D$; (iii) conforms to a receiver-projected action signature,
an implementation-declared effect in $\Omega$, and the issue-time argument digest in $B$;
(iv) enters action code for its \emph{first} target invocation while every active member, policy,
executor, entitlement, and query-witness fingerprint and the row state of every target in $A$
equal their issue-time values, and thereafter, until commit, no writer other than
the action invocations of this same execution changes state in
$\operatorname{dom}(\Delta\!\downarrow_A)$;
and (v) is the sole consumer of $(\iota,n)$, consumed at an acceptance instant in $\mathcal{J}$. Hence
a model-generated continuation can select only within the authority envelope established by
query evaluation and controller projection; it cannot amplify targets, operations, effects,
identity, world, provenance, or lifetime at the receiver-bound action-entry boundary.

\textit{Proof sketch.} (i) is MAC verification with the issuer, world, and execution-context
checks; (ii) is Proposition~3; (iii) follows from the grant premise of Eq.~\eqref{eq:qrc-auth}
and fresh materialization from the same immutable argument plan at each target; (iv) is
Proposition~4; (v) is Proposition~5. All are established before the first target enters action
code under one stable view. Under the grant-provider contract in Section~\ref{sec:trust},
the fingerprint equalities in (iv) also establish that the action and effect in (iii)
remain authorized for the original request. Equality in (iv) is equality of the recorded
representations. Inferring that the query still selects a target additionally requires the
read-semantics premise of Theorem~2.

\textbf{Corollary 1 (repository-effect confinement).}
Call an action implementation \emph{receiver-confined} when every repository object it mutates is
the receiver supplied by the dispatcher and an update preserves that receiver's identifier,
and \emph{effect-sound} when its realized repository
effect belongs to its declared effect set. For receiver-confined, effect-sound implementations,
every mutation produced by an invocation admitted under $C$ addresses an object in $A\subseteq D$
and realizes an effect in the grant recorded by $\Omega$.

\textit{Justification.} Theorem~1 fixes the receiver instances supplied at action entry and the
permitted effect class. Receiver confinement ensures that every repository mutation addresses
one of those instances, and effect soundness ensures that it realizes a granted effect.
The non-versioned, receiver-local update and delete paths used by
the QRC evaluation satisfy these assumptions, as do the three evaluated domain actions.
The standard update rejects assignments to the identity field before repository writes or
mutation-event publication, keeping the authenticated target and the event identity aligned.
Creation establishes a new object rather than mutating an object in $A$; versioned singleton
updates additionally insert a successor and retire prior current versions. Those operations
retain their existing platform semantics and require authority over their complete effect
scope rather than an application of this receiver-local corollary.

\subsection{Continuation guarantees and enforcement}
\label{sec:scope}

Theorem~1 associates each continuation with the runtime state that justifies its authority.
Explicit dependency ownership supports selective revalidation as the ontology and data evolve.

\textbf{Selected-state freshness and its stronger variants.} $\Delta$ contains schema
fingerprints for every field and edge the query referenced and row-state fingerprints for the
selected objects $D$ of receiver type $T$. A dispatch activates the full-row keys for its chosen
targets $A$. QRC therefore certifies that the receivers handed to the action are objects the
accepted query denoted and that, when the batch first enters action code, every target is still in
the state in which the query observed it---the semantics of an ETag-conditioned update
\cite{odata-protocol-2020}, generalized from one addressed resource to a query-bounded target set
and combined with the ontology dependencies that produced it. Let $D'=[\![Q]\!]_{S'}$ be the denotation $Q$ would have under the
dispatch-time view. Three guarantees of increasing strength should be kept apart; each is
stated as a conjunction with the one before it, so that G3 implies G2 and G2 implies G1:
\begin{itemize}
\item[(G1)] \emph{historical selection}: $A\subseteq D$, and every target is in its
issue-time state at the batch's first action entry; afterwards only this batch's own action
invocations may change a selected row before commit. This is what Theorem~1 establishes and
what the default selected-row profile enforces. The stronger reading---each target in its issue-time
state at its \emph{own} action entry---holds in addition when the action is receiver-confined
(Corollary~1), since then the batch's earlier invocations wrote only their own receivers.
\item[(G2)] \emph{target re-satisfaction}: G1 and $A\subseteq D'$ in the stable view
  immediately before the batch's first action invocation.
\item[(G3)] \emph{denotation re-satisfaction}: G2 and $D=D'$, i.e., the query result as a whole
is unchanged.
\end{itemize}
G2 uses one batch-entry view. Even a receiver-confined action can update a row that a later
target reaches through a link: if one target is another's manager, updating the manager's
status can change the second target's predicate before its own invocation. Corollary~1
preserves the later target's own row in this case, not all rows read by its predicate.
A workflow requiring membership before \emph{each} invocation needs per-target revalidation
or an action contract preserving those predicates. Target coverage is a separate choice:
operating on every current match requires $A=D=D'$, whereas G3 permits $A\subseteq D$.
The default G1 profile binds selected root rows. A changed laptop return date can therefore
leave an employee capability valid, even though that employee no longer satisfies the original
navigation predicate. The functional-witness profile derives and locks the navigated state
needed by $\mathcal{Q}_F$ and establishes G2 by Theorem~2. General queries can instead be
re-evaluated under the dispatch snapshot, with $A\subseteq D'$ checked before execution.
The SQLite business study evaluates that alternative within the same write transaction.
Table~\ref{tab:business-contracts} maps the three continuation guarantees to the dependency
policies that realize them.

\begin{table}[!htbp]
\centering
\caption{Dependency policies for business continuations}
\label{tab:business-contracts}
\small
\begin{tabular}{@{}>{\raggedright\arraybackslash}p{3.4cm}>{\raggedright\arraybackslash}p{3.0cm}>{\raggedright\arraybackslash}p{5.0cm}@{}}
\toprule
Workflow requirement & Required guarantee & Realization \\
\midrule
Apply a previously approved target set while its records are unchanged
 & Historical selection (G1) & Target rows and referenced schema, grant, and policy keys \\
Continue recovery while each selected device remains unreturned
 & G1 and current membership (G2) & Functional witnesses for $\mathcal{Q}_F$, or query re-evaluation \\
Keep the approved query result equal to the current result
 & G2 and result equality (G3) & Re-evaluation with $D=D'$ and a suitable concurrent-writer protocol \\
\bottomrule
\end{tabular}
\end{table}

\textbf{Profile selection and business policy.} A business requirement and a witness mechanism
are distinct: G2 is a requirement, whereas functional witnesses are one way to meet it.
There are two useful approval policies within G2. A \emph{current-membership approval} accepts
the invoked targets when they still satisfy the selecting predicate and retain their approved
row states. A \emph{recorded-evidence approval} additionally compares the query-observed
values and links for all approved objects $D$ at issuance and batch entry. If an observed
value changes and is restored before entry, it satisfies this equality. The approved query
determines the evidence: a check on a supplier's rating binds that rating; a check on
its certificate also binds the traversed link and referenced certificate field. The trusted
controller selects the query and decides which evidence the action requires.
A path-sensitive footprint records syntactic
read paths. The Java functional profile records a schema-guided superset; repeated receiver
types can enlarge that footprint and thereby narrow the set of accepted continuations.
In the procurement study (Section~\ref{sec:contract-evolution}), a supplier's rating rises
from 80 to 90 while the order and the eligibility threshold of 70 are unchanged. If the
instruction is to mark the order as reviewed provided the supplier still qualifies, both
membership controllers admit the update. If the reviewer signed off \emph{on the recorded
rating}, both evidence controllers require renewal: 90 meets the threshold but is not the
rating the reviewer saw. When eligibility also depends on the supplier's certificate, the
compiler adds the certificate link and validity field to the witness footprint; the JDBC
evidence controller requires the corresponding projection update. Evidence for non-invoked members
also remains active when the footprint covers the approved batch.
Section~\ref{sec:footprint-properties} defines the footprints;
Section~\ref{sec:footprint-precision} measures their difference.

An action that requires the original observations sets its own evidence minimum.
The Java implementation checks the live descriptor even when several issuer profiles serve
the same receiver. Upgrading the descriptor changes its fingerprint and retires earlier grants;
a membership-only issuer cannot mint a replacement for a strict action.
Let $g_{\min}^{S}(T,a)$ be the minimum guarantee declared by the trusted action descriptor,
and let $g(\chi)$ be the guarantee attested by the configured service after query classification.
The issuance service enforces
\[
 \forall a\in\operatorname{dom}(\Omega),\qquad
 \mathbf{s}_\chi\succeq\mathbf{m}_{S}(T,a),\qquad
 G3\succeq G2\succeq G1.
\]
The descriptor's \texttt{minimumContinuity} is implementation-owned and included in its
fingerprint. The service fixes \texttt{WitnessMode} at construction and authenticates it in the
capability; a caller cannot elevate the recorded level by editing the payload. Dispatch checks
the current descriptor's requirement before nonce consumption and carries the established level
in the invocation-bound permit. The action entry checks that permit again. Ordinary direct
and registry calls cannot enter an action with a declared requirement without such a permit.
Descriptors with neither requirement retain their direct-call semantics.
G2-protected actions declare G2 in their descriptor and require the corresponding entry permit.

\paragraph{Independent evidence requirement}
The descriptor for $(T,a)$ may declare $e_{\min}\in\{0,1\}$: 1 requires
equality of the query-observed footprint $P_Q(D,S_0)$ defined in
Section~\ref{sec:footprint-properties}. Let $e(\chi)=1$ for a trusted profile that records
and validates this footprint or a superset, and 0 otherwise. Admission requires
$g(\chi)\succeq g_{\min}$ \emph{and} $e(\chi)\ge e_{\min}$. Issuance compares both
coordinates; dispatch authenticates the issuer and profile and verifies the descriptor against the authenticated issue-time record. In Java the registry lock
holds that descriptor stable through execution; the action tests both coordinates on the
invocation-bound permit. The independent prototype instead carries and checks the descriptor
revision on its permit. A revision retires outstanding grants, including those whose profile
would meet the new requirement. Thus a G2 recheck credential cannot authorize a strict
action, even when an independently keyed recheck issuer remains active.

\textbf{Strict-entry property.} Suppose the approved query and grant projection are trusted,
descriptor revisions bind their meanings, recorded values identify the intended rows and
columns faithfully, and the verified view is stable until the batch first enters action code.
If the live descriptor requires $e_{\min}=1$, every admitted batch then agrees with $S_0$ on
$P_Q(D,S_0)$. The grant and revision checks preclude substituting another action or an earlier
policy. The componentwise minimum excludes membership-only profiles. Every remaining profile
captures at least $P_Q$ and compares its values before entry. Equality therefore holds even
when a different policy could have admitted the same targets by rechecking membership alone.
Section~\ref{sec:prototype} tests the two coordinates separately.

\paragraph{Permit scope in the Java runtime}
A permit binds one action and its invocation targets. A nested protected action requires
a permit of its own. SQLite supports transactional query rechecking; PostgreSQL supports
the evaluated witness profiles. Descriptor persistence, nested calls, cached entry points,
and failure semantics are documented in Supplementary Sections~S5 and S12.

\textbf{No-downgrade property.} For a protected registered action, every permitted entry through
the QRC and ordinary registry APIs meets the live minimum in both policy coordinates. Issuance
checks all projected actions, dispatch checks the requested action under the registry lock,
and the entry permit checks the same requirement and consumes one target authorization.
Changing the profile breaks authentication; raising a minimum either fails the strength test
or changes the recorded action fingerprint. Direct entry has no permit and is refused.
The property holds at the registered action-entry boundary under the trusted-host assumptions
of Section~\ref{sec:trust}; action-body behavior is governed by the receiver-confinement and
effect-soundness conditions of Corollary~1.

G3 concerns changes outside the selected objects as well: a newly matching employee can enlarge
$D'$ while leaving every captured row unchanged. Ordered, bounded, aggregate, and absence-based
queries likewise require their own predicate or read-set treatment. Re-evaluation defines
membership in the dispatch snapshot; preserving the navigated state through action entry
additionally requires locks or serialized writers covering that state. Serializable execution
provides a serializable outcome with retryable transaction failures, which is distinct from
blocking every concurrent writer. The evaluated witness profiles cover single-valued
row-local navigation; general G3 support remains an extension.

\textbf{Creation.} A \texttt{create} terminal has no selecting denotation. Collection creation is
governed by the collection receiver's projected create contract and application authorization
rather than by a QRC.

\textbf{Deployment topology.} The issuer is bound to one world. The evaluated world and
registry monitors coordinate one runtime process; multiple processes sharing a world
require coordinated configuration versions. Supplementary Sections~S5 and S13 detail
failure handling and recovery of single-use grants.

\textbf{Trusted implementations.} Corollary~1 composes Theorem~1 with receiver-confined,
effect-sound action implementations. QRC protects the action-entry boundary; the body of a
host-language action is trusted deployment code.

\section{Ontology Program Execution}
\label{sec:execution}

\subsection{Typed selection and execution}
\label{sec:substrate}
\label{sec:lazy}

General VirtualSet expressions compose navigation, projection, grouping, and materialization
through the receiver-transition rules of Section~\ref{sec:type-continuity}. QRC issuance
selects the read-only, single-root fragment whose result retains the root entity domain
and whose complete selection lowers to SQL. Within that fragment, $\mathcal{Q}_F$ admits
row-local scalar predicates and single-valued navigation for automatic G2 witnesses.

The complete-AST validator establishes member ownership, signature compatibility, and callback
types before repository access. \texttt{GCPToSQL} fuses translatable predicates and uses
schema-guided subqueries for navigation. Missing single-edge targets yield \texttt{NULL};
identifier-existence tests distinguish them from present targets with null fields. Ordinary
reads may evaluate a residual tail through \texttt{StreamSetValue}, with a default 10,000-row
materialization guard. QRC issuance requires full lowering so that selection and dependency
capture share a transaction-stable repository view.

A model-facing controller supplies candidate expressions and trusted operation requests.
Generation strategy remains independent of the validator and authority compiler: accepted
selection supplies the receiver and denotation, the controller attenuates permitted operations,
and the action runtime consumes the resulting authority through its receiver-bound entry path.

\subsection{QRC implementation and stable dispatch}
\label{sec:qrc-impl}

\textbf{Integration with the existing runtime.} The implementation connects VirtualSet's query
and action interfaces through a result-specific authority layer. Its selection adapter follows the same
principal-scoped path as \texttt{queryIds}: read authorization precedes typed evaluation, the
accepted expression must evaluate to an entity collection, its runtime schema must match the
inferred receiver, and the result is an immutable identifier list. Its action adapter resolves the
same receiver-qualified \texttt{ActionDescriptor} and preserves the permission and authority
checks, entity loading, action body, session, and event semantics of \texttt{executeAction}. QRC
issuance stores the additional receiver, denotation, grant, argument,
and dependency evidence; dispatch validates that evidence and creates an invocation-bound permit
before entering the receiver-bound action adapter. This integration enforces query-derived
authority while preserving the query evaluator and registered action semantics.

\textbf{Fingerprints.} Table~\ref{tab:dependency-map} connects each dependency source to its
captured key. A selected-row key encodes all non-null stored column names, runtime types, and
values in canonical order. Full-row keys are recorded and captured for $D$ but compared only
for invoked targets $A$. Thus a populated target-column change invalidates G1, while an
unpopulated nullable-column addition does not.

For a scalar schema key, Java fingerprints the owning type, field name and mapped SQL
column type. In a procurement regression, changing the supplier rating from \texttt{Int}
to \texttt{Long} preserved its \texttt{INTEGER} key, the three selected identifiers and the
original approval. Changing the declaration to \texttt{String} changed the mapped type and
blocked the action. Advancing the executor version during the \texttt{Int}-to-\texttt{Long}
change also blocked it, this time with the field key unchanged. These checks distinguish
a change the schema key detects from one the deployment explicitly versions.

Functional witnesses instantiate the query-referenced scalar fields and foreign-key columns
over the schema-guided closure. They acquire row locks while comparing field-level fingerprints,
admitting changes to columns outside that footprint. Witness keys derived for all of $D$ remain active and may therefore reject
a change outside $A$ when it affects recorded evidence. Unrelated evolution is defined relative
to the complete active footprint $\Delta\!\downarrow_A$. Action fingerprints include the resolved
descriptor and minimum guarantee; trusted provider versions bind policy, entitlements, and
executor semantics. An executor upgrade changes its deployment digest.

\begin{table}[!htbp]
\centering
\caption{Fingerprint keys for query and grant dependencies}
\label{tab:dependency-map}
\scriptsize
\setlength{\tabcolsep}{3pt}
\begin{tabular}{>{\raggedright\arraybackslash}p{3.3cm}>{\raggedright\arraybackslash}p{3.7cm}>{\raggedright\arraybackslash}p{5.2cm}}
\toprule
Dependency source & Key & Fingerprint material / invalidating change \\
\midrule
Field read by $Q$, under its owning receiver & $\textsc{Field}(T,f)$ & owner, field name, mapped SQL type; field removed or renamed, or mapped type changed \\
Edge navigated by $Q$, under its owning receiver & $\textsc{Edge}(T,e)$ & owner, name, target type, cardinality, key column; any change to the edge \\
State of each selected object & $\textsc{EntityState}(T,id)$ & non-null column name/type/value encoding; recorded for $D$, compared for $A$. A changed stored value or deletion invalidates an invoked target; an unpopulated nullable column or same-value write does not \\
Query-observed value of a navigated object & $\textsc{EntityFieldState}(T,id,f)$ & receiver, identifier, storage column, runtime type, and value; changes to the referenced scalar field or navigation foreign key, deletion of the row, or loss of the column; writes to other columns do not invalidate the witness \\
Descriptor of each granted action & $\textsc{Action}(T,a)$ & resolved exact-or-wildcard key, entry revision, signature, source or endpoint, parameters, timeout, effects, continuity and evidence minima; re-registration of that descriptor only \\
Policy that projected $\Omega$ & $\textsc{Policy}(T,\pi)$ & provider fingerprint for the policy and its grant-relevant inputs, together with the entitlement key below \\
Entitlements of $p$ on $T$ & $\textsc{Principal}(T,p)$ & provider digest of the principal's permissions on the receiver \\
Evaluator, operator semantics, translation & $\textsc{Executor}(T,x)$ & deployment version of the executor for the receiver; any evaluator upgrade \\
User-defined and host function bodies & none & excluded from the issuable sublanguage by the static check, before evaluation \\
Storage mapping of members & via \textsc{Field}/\textsc{Edge} & identity mapping on member names; a mapping change is a member change \\
\bottomrule
\end{tabular}
\end{table}

\textbf{Issuance.} The service receives a query, principal, requested actions, and arguments.
Within the stable issuance transaction it validates types, checks the query grammar, derives
the receiver and schema keys, evaluates selected identifiers, projects grants, and captures
state. The admissibility check uses an explicit syntax and call whitelist: immutable aliases,
lexical blocks, scalar operators, field reads, declared navigation, and read-only collection
operators. Functions occur as direct element callbacks. Every call must resolve to a declared
edge or a whitelisted operator on a known ontology receiver; actions, free calls, structural
object methods, mutable assignments, and unrecognized syntax produce a static rejection.
The evaluated result must then be a completely SQL-lowered lazy selection of the root type.

These checks separate three properties: typing establishes member and argument compatibility;
admissibility establishes the read-only query fragment; SQL lowering establishes executable
selection semantics. The end-to-end regressions include a structural-object method wrapping an
action and verify zero action entries and unchanged rows. A top-level collection alias is
issuable; a block-local navigation alias is analyzable but currently reaches the stream fallback
and is rejected at the lowering check.

\textbf{Dependency extraction.} After type validation, Algorithm~\ref{alg:dependency-extraction}
derives the schema keys $K_Q$ and checks syntactic admissibility in one pass over the
accepted AST. The issuer subsequently captures the fingerprints of these keys. The traversal
propagates a \emph{receiver flow}---$\mathbf{none}$, $\mathbf{entity}\,T$, or
$\mathbf{coll}\,T$---through the expression under an environment $\Gamma_\flat$ that binds
lambda parameters and block-local aliases to flows. A member selected on a known flow is
recorded under the flow's type: a field yields $\textsc{Field}(T,f)$ and ends the flow; an edge
yields $\textsc{Edge}(T,e)$ and continues the flow at the edge's target type with the edge's
cardinality. A collection operator that applies a callback to elements (\texttt{filter},
\texttt{map}, \texttt{order\_by}, \texttt{any}, \texttt{sum}, \ldots) binds the lambda
parameter to $\mathbf{entity}\,T$ before the body is traversed, which is what attributes
\texttt{e.age} in the running example to \texttt{Employee} and \texttt{returnDate} to
\texttt{Laptop} even if both types declared a field of the same name. Collection-preserving
operators return $\mathbf{coll}\,T$; \texttt{map} returns a collection of the callback's flow;
element operators return $\mathbf{entity}\,T$. A member that resolves to a registered action or
effectful operator, and any call whose callee is not a member of a receiver flow, is a violation;
the query is admissible iff there is no violation, exactly one collection root was reached, and
the program's flow is $\mathbf{coll}$ of that root. The keys the algorithm never records are
those Table~\ref{tab:dependency-map} lists as needing none.

\begin{algorithm}[!tb]
\caption{Receiver-flow extraction over the read-only query core.}
\label{alg:dependency-extraction}
\footnotesize
\begin{algorithmic}[1]
\Function{Extract}{$e,\Gamma_\flat$}
  \If{$e$ is a bound variable $x$}
    \Return $\Gamma_\flat(x)$
  \ElsIf{$e$ is a collection root of type $T$}
    \State record root $T$; \Return $\mathbf{coll}\,T$
  \ElsIf{$e$ is an immutable alias \texttt{let} $x=e_0$}
    \State $\Gamma_\flat[x]\gets\Call{Extract}{e_0,\Gamma_\flat}$; \Return $\Gamma_\flat[x]$
  \ElsIf{$e$ is a lexical block}
    \State extract statements in order under a copy of $\Gamma_\flat$
    \State \Return the final statement's flow
  \ElsIf{$e=e_0.m$ or $e=e_0.m(a_1,\ldots,a_k)$}
    \State $\phi\gets\Call{Extract}{e_0,\Gamma_\flat}$
    \State require a known receiver for a call, and an edge or whitelisted read operator
    \State reject registered actions and effectful members
    \If{$m$ is a field of $\operatorname{type}(\phi)$}
      \State record its owner-qualified field key; \Return $\mathbf{none}$
    \ElsIf{$m$ is a declared edge}
      \State record its owner-qualified edge key; derive its target flow $\phi'$
    \EndIf
    \ForAll{$a_i$}
      \State extract $a_i$; a direct element callback receives $\mathbf{entity}\,\operatorname{type}(\phi)$
    \EndFor
    \State \Return $\phi'$ for navigation; host collection flow for preserving operators;
      callback collection flow for mapping; host entity flow for element operators;
      $\mathbf{none}$ for scalar operators
  \ElsIf{$e$ is a scalar literal or primitive scalar expression}
    \State extract operands; \Return $\mathbf{none}$
  \ElsIf{$e$ is a program or statement wrapper}
    \State extract children in order; \Return the final child's flow
  \Else
    \State record an unsupported-syntax violation; \Return $\mathbf{none}$
  \EndIf
\EndFunction
\Statex Accept iff there are no violations, one root type, and the final flow is its collection.
\Statex Calls outside a resolved ontology receiver and functions outside direct callbacks are rejected.
\end{algorithmic}
\end{algorithm}

\paragraph{Schema coverage and functional data witnesses}
\label{sec:footprint-properties}
The derivation has two stages: schema-occurrence coverage fixes the ontology declarations on
which the query depends, and data-witness closure fixes the row values needed to preserve
membership for a functional-navigation fragment.

\textbf{Definition (flow-transparent core).} Let $\mathcal{Q}_0$ contain programs built from a
typed ontology collection root, immutable identifier aliases, lexical blocks, scalar literals
and operators, receiver-qualified fields and declared navigation, and read-only collection
operators with the flow rules of Algorithm~\ref{alg:dependency-extraction}. Lambdas occur as
direct element callbacks; their parameters inherit the host element type. Receiver-valued
expressions have a statically propagated entity type and cardinality. Mutable assignments,
conditional receiver merges, object construction, dynamic members, and opaque calls are excluded.
The static whitelist implements these constructors; complete SQL lowering further restricts
the issuable subset, whose final value is a collection of the root type.

Write $\operatorname{Occ}(Q)$ for the set of owner-qualified field and edge keys named
by syntactic member occurrences of $Q$, including callback bodies. Primitive operators and
collection-operator meanings are fixed by the executor version.

\textbf{Lemma 1 (schema-occurrence coverage).} For a well-typed $Q\in\mathcal{Q}_0$,
Algorithm~\ref{alg:dependency-extraction} derives exactly $\operatorname{Occ}(Q)$:
$K_Q=\operatorname{Occ}(Q)$, with $K_Q=\operatorname{dom}(\Delta_Q)$ at issuance.
Alpha-renaming a bound variable, or introducing a fresh
lexical alias for a receiver path, preserves this set.

\textit{Proof.} Induct over the core syntax, maintaining that every receiver-valued expression
has its resolved type and cardinality. Roots establish collection flows, and variables retrieve
their lexical bindings. A field records its owner's field key and terminates receiver flow.
An edge records its owner's edge key and propagates its declared target flow. Element callbacks
extend a copied environment with the host element type. Aliases retain the derived flow; blocks
restore outer bindings on exit and return their final statement's flow. Primitive scalar
expressions visit all operands and return scalar flow. Collection-preserving operators retain
the host flow, element-returning operators retain its entity type, and mapping propagates the
callback flow. Thus every schema occurrence is recorded with its owner, and each recorded key
comes from such an occurrence. Renaming and fresh alias introduction preserve these owners.
$\square$

For $n$ AST nodes, at most $s\ge1$ live bindings, depth $h$, and $k$ keys, environment copying
gives $O(ns+k)$ expected time and $O(hs+k)$ auxiliary space, assuming constant-cost schema-map
operations. The service obtains receiver, admissibility, and schema keys from one traversal.
Seven analyzer regressions cover name collisions, navigation, alias and alpha equivalence,
composition, scalar shadowing, and restoration after nested blocks.

For a captured footprint $F$, let
$\mathcal{V}(F,S_0)=\{S:\forall d\in F,\ d\in\operatorname{dom}(S)\land S[d]=S_0[d]\}$.
Adding witnesses refines its valid-view set: $F\subseteq F'$ implies
$\mathcal{V}(F',S_0)\subseteq\mathcal{V}(F,S_0)$, because every equality for $F$ also occurs in
$F'$. The next result identifies a useful sufficient footprint.

\textbf{Functional-navigation closure.} Let $\mathcal{Q}_F\subseteq\mathcal{Q}_0$ contain
root selections composed of \texttt{filter} predicates whose scalar reads follow finite paths
of single-valued edges from the current element. Predicates use deterministic scalar operators
and immutable aliases. Global collection reads, aggregation, ranking, and result truncation
are outside $\mathcal{Q}_F$. Let $E_Q$ contain $(T,e,T')$ whenever
$\textsc{Edge}(T,e)\in K_Q$ and the issue-time declaration has target type $T'$.
Starting with $K_0=\{(T,id):id\in D\}$, compute
\[
\begin{aligned}
 K_{j+1}&=K_j\cup
 \{(T',\operatorname{fk}_{S_0}(T,id,e)):\ (T,id)\in K_j,\\
 &\hspace{3.5em}(T,e,T')\in E_Q,\
       \operatorname{fk}_{S_0}(T,id,e)\ne\mathrm{null}\},\\
 K^*&=\bigcup_{j\ge0}K_j.
\end{aligned}
\]
Here $\operatorname{fk}_{S_0}$ reads the stored target identifier. A visited set deduplicates
shared endpoints and cycles. The implementation reads each visited row once during closure
construction in the issue transaction; a missing non-null endpoint or the dependency bound
stops issuance. With $v$ visited rows, $e$ traversed edge instances, and $r$ scalar-field
occurrences instantiated over those rows, traversal uses $O(v)$ row lookups and
$O(v+e+r)$ expected local work and auxiliary space. The implementation
also sorts the $|D|$ initial identifiers and the $k=|K_Q|$ schema keys, adding
$O(|D|\log |D|+k\log k)$ work before traversal; fingerprint capture follows. The closure is a schema-guided overapproximation: each edge referenced for a type is
followed on every reached row of that type. For a reached row define its captured-field set
\[
\begin{aligned}
\mathcal{O}_Q(T',id)={}&
 \{\textsc{EntityFieldState}(T',id,f):\textsc{Field}(T',f)\in K_Q\}\\
 &\cup\{\textsc{EntityFieldState}(T',id,\operatorname{fk}(e)):
       \textsc{Edge}(T',e)\in K_Q\}.
\end{aligned}
\]
The resulting field-sensitive profile binds
\[
 F_F=K_Q\cup K_{\Omega}\cup
     \{\textsc{EntityState}(T,id):id\in D\}\cup
     \bigcup_{(T',id)\in K^*}\mathcal{O}_Q(T',id).
\]
Here $K_{\Omega}=\operatorname{dom}(\operatorname{Deps}_{S_0}(\Omega,p))$ is the set
of grant-related keys. For the functional-witness profile, $F_F$ is the domain of
the signed fingerprint map $\Delta$.

\textbf{Evidence precision.} Let $P_Q(D,S_0)$ contain the field-state keys obtained by
instantiating each syntactic scalar-read path and its foreign-key prefixes on every root in
$D$. Paths stop at their syntactic endpoint or a null link; both sides of a scalar condition
are included. This path-sensitive footprint is a syntactic dependency set, not a minimum
semantic explanation of a true predicate. Write
$W_Q(D,S_0)=\bigcup_{(T',id)\in K^*}\mathcal{O}_Q(T',id)$ for the Java runtime's closure.
Every finite query path is included in the closure, so
\[
 P_Q(D,S_0)\subseteq W_Q(D,S_0),\qquad
 \mathcal{V}(W_Q,S_0)\subseteq\mathcal{V}(P_Q,S_0).
\]
The first inclusion follows by induction along each path; the second follows from footprint
monotonicity. These relations separate sufficient evidence from precision. For
\texttt{employees.filter(e -> e.manager().age >= 65)}, with an Employee-valued
\texttt{manager}, the closure can include the manager's manager although the predicate stops
after one hop. Changing that further object's captured age can invalidate $W_Q$ while leaving
$P_Q$ and target membership unchanged. Section~\ref{sec:footprint-precision} measures this
effect through complete dispatches. Theorem~2 establishes sufficiency for $W_Q$.

\textbf{Finite-path fragment.} In the common fragment of the Java and SQLite path issuers,
a query has the following syntax (up to binder names and parentheses):
\[
\begin{aligned}
 Q &::= R\,F^+, & F &::= \texttt{.filter(}x\mathbin{\to}b\texttt{)},\\
 b &::= p\,\theta\,n \mid b\land b, &
 p &::= x\,(\texttt{.}u\texttt{()})^*\,\texttt{.}f .
\end{aligned}
\]
$R$ is a mapped root collection, $u$ a declared single-valued edge, $f$ an integer column,
$n$ an integer literal and $\theta\in\{=,<,\le,>,\ge\}$. The SQLite comparator instantiates
this grammar over \texttt{items}; Java resolves owners on the accepted AST and also supports
textual equality with a string literal. For each selected root, the issuer records
$(\text{owner type},\text{row id},\text{foreign-key column})$ at each step and the scalar key
at a reached leaf. A null link contributes its own key and stops that path. Shared keys occur
once, and revisiting a row consumes a step of finite syntax. Aggregation and result truncation
are outside this profile.

\textbf{Path sufficiency.} For this fragment, if the mappings are stable, non-null links
resolve, the referenced field and operator meanings are unchanged, and the target rows
and $P_Q(D,S_0)$ agree with their issue-time values through
first action entry, every invoked target still satisfies $Q$ at batch entry.
Fix such a target. Its first preserved link fixes the next row or null; induction along the
path preserves every subsequent endpoint. The terminal comparison therefore has the same
operands and truth value under the same null semantics. Conjunction and successive filters
preserve membership of that target. This argument uses the path keys directly: sufficiency
of the larger closure would not, by itself, imply sufficiency of its subset.

\begin{samepage}
\textbf{Theorem 2 (functional-witness sufficiency).} Let $Q\in\mathcal{Q}_F$ select $D$ in
the stable issue view $S_0$, with every non-null navigated identifier resolving to a row.
Let $\emptyset\ne A\subseteq D$ be the invocation targets. Assume unchanged executor and
referenced schema meanings, faithful row and field fingerprints, and successful validation of
$F_F\!\downarrow_A$ under the stable locking region of Algorithm~\ref{alg:qrc-dispatch}.
At the batch's first action entry, $A\subseteq[\![Q]\!]_{S'}$: every invoked target still
satisfies $Q$, and the invocation has G2 as well as G1. No membership conclusion is required
for $D\setminus A$.
\par
\end{samepage}

\textit{Proof.} Fix $d\in A$. Its active full-row key establishes that the root still exists
and retains its issue-time values, including when the predicate is constant and has no field
witnesses. The closure seeded by $D$ contains every row reachable along a predicate path from
$d$, since $A\subseteq D$. Restriction to $A$ removes only non-target root full-row keys;
it removes none of the scalar or foreign-key witnesses on those paths.
Induct along each predicate navigation path. A null foreign key remains null; a non-null
foreign key is preserved by its \textsc{EntityFieldState} key and leads to the same target
identifier. Every scalar read on the reached target has its own field-state key, and every
subsequent navigation has a key for its foreign-key column. Lemma~1 accounts for the declarations
interpreting every step. All scalar leaves therefore retain their issue-time values.
Deterministic scalar composition preserves each predicate result, and a sequence of filters
preserves membership of $d$. Since $d$ was arbitrary, every member of $A$ still satisfies $Q$.
The stable dispatch region extends these equalities to first action entry. Together with
Theorem~1, this gives G2.
$\square$

For example, a constant filter may select $D=\{d_1,d_2\}$ and a continuation target only
$A=\{d_1\}$. Deleting $d_2$ before dispatch changes an inactive full-row key and need not
prevent the invocation on $d_1$. This illustrates G2's target-specific guarantee for $A$.
Conversely, deletion of $d_1$ changes an active key and prevents action entry.

\textbf{Why membership does not determine evidence equality.}
Fix a query selecting an order when its supplier's rating is at least 70. Two dispatch
views agree on the target row, grant and current answer. In the first the rating remains
80; in the second it is 90. An approval requiring the recorded rating to match admits
the first and refuses the second. Any rule observing only the query identity, target
rows, membership and unchanged grant receives identical inputs and must return the
same verdict. Recording the rating under its actual owner supplies the missing distinction.

\textbf{Theorem 3 (end-to-end receiver and authority continuity).}
Let $Q$ be an accepted, issuable query under $S_0$, with final flow $\mathbf{coll}\,T$,
denotation $D$, and receiver-qualified dependencies $\Delta_Q$. Let a trusted controller approve
$\mathcal{R}_{\mathrm{req}}$ and project $\Omega$, and let the issuer construct $C$ from this
query judgment and projection. Every later invocation $I$ admitted under $C$ satisfies:
(i) each member occurrence in $Q$ was validated against its inferred receiver, and each query
dependency is attributed to its owning receiver; (ii) the action receiver is $T$, its targets
form a nonempty $A\subseteq D$, and its operation, arguments, effect, principal, world, and
acceptance instant satisfy the authenticated controller/issuer envelope; and (iii) its active
captured-state and referenced schema, action, policy, executor, and entitlement dependencies
agree with $S_0$ at first action entry; and (iv) the verified continuation level meets the
current action descriptor's continuity minimum. If the descriptor also enforces the evidence
coordinate, the strict-entry property in Section~\ref{sec:scope} additionally applies
when $e_{\min}=1$. With functional witnesses for $Q\in\mathcal{Q}_F$ and the
semantic and fingerprint assumptions of Theorem~2,
$A\subseteq[\![Q]\!]_{S'}$ also holds.

If $Q$ occurs inside a separately validated complete program $P$, Proposition~1 additionally
establishes member legality throughout that $P$ at its validation view. In decoupled operation,
$I$ may be formed after issuance as an independent request; capability verification establishes
its authority at dispatch. The invariant thus composes query typing, trusted authority
projection, and later dispatch across their respective inputs and runtime views.

\textit{Proof.} Apply Proposition~1 to $Q$ and Proposition~2 to its receiver transitions.
Lemma~1 records each field and edge under its propagated owner, establishing (i).
Issuance authenticates the evaluated $T,D,\Delta_Q$ together with the trusted projection's
grants and arguments and the issuer's identity, world, and lifetime. Theorem~1 then establishes
(ii) and (iii) for any admitted $I$, regardless of when that invocation was constructed.
The descriptor and permit checks establish (iv). Theorem~2 supplies target membership under
the functional profile; protected rechecking establishes it by evaluating $Q_r$ in the stable
dispatch view. Applying Proposition~1
separately to $P$ gives the additional whole-program legality statement. $\square$

The service selects the closure profile through \texttt{FUNCTIONAL\_NAVIGATION}; its classifier accepts
row-local filters and single-valued navigation before closure construction. Selected-row G1
remains the compatibility default and cannot issue authority for an action requiring G2.
The \texttt{TYPED\_PATH} mode uses the finite-path classifier above and rejects selections
outside that fragment. Field-sensitive binding admits changes to unobserved witness values, while
insertions outside the closure preserve existing targets' membership.
Section~\ref{sec:business-workflows} measures both effects against query re-evaluation.

\textbf{Worked example.} The filter defining \texttt{at\_risk} in Listing~\ref{lst:risk-query} yields four
schema keys: \texttt{Employee.age}, \texttt{Employee.assignedLaptop}, \texttt{Laptop.id},
and \texttt{Laptop.returnDate}, each qualified by its owner and kind.
The laptop value used only for the displayed total is not a dependency of this filter. Issuance adds selected
employee rows and action, policy, principal, and executor dependencies.
If the selected employee returns the laptop before dispatch, the employee row is unchanged:
G1 admits the update, but the functional witness detects the change to the linked laptop's
return date. Adding an unused nullable field to the schema invalidates neither contract.
Supplementary Material, Section~S18 gives the revision-by-revision decisions.

\textbf{Dispatch.} Under the lock order of Algorithm~\ref{alg:qrc-dispatch}, the
dispatcher validates the authenticated argument tree and materializes a fresh value for
each target. Only after verification and nonce consumption does it create the permit
accepted by the action adapter. Per-target savepoints cover local updates and synchronous
events. Section~\ref{sec:trust} specifies the retry boundary; Supplementary Sections~S1
and~S5 give the backend tests, payload bounds and failure conditions.

\textbf{Issuable fragment.} Ordinary reads may take either path of Section~\ref{sec:lazy}, and
Section~\ref{sec:evaluation} shows that they agree. The implementation \emph{issues} capabilities
only for selecting chains that \texttt{GCPToSQL} lowers completely. The formal contract also admits an
evaluation path that yields the same receiver, denotation, and dependency keys. Supporting the
current stream path at issuance additionally requires purity certification for runtime-executed
lambda bodies.

\subsection{Receiver-bound CRUD and domain actions}
\label{sec:crud}

VirtualSet provides three opt-in persistent operations. A typed collection may declare
\texttt{create(...)}, whereas an existing entity may declare \texttt{update(...)} and
\texttt{delete()}. Higher-level domain actions are registered for a receiver type through an
\texttt{ActionDescriptor}. Before dispatch, a call is checked against that receiver's member
closure and declared signature. On the query-derived route, existing-entity update, delete, and
domain-action calls are additionally checked against the QRC derived from the selecting query;
collection creation remains governed by
the collection receiver's projected create contract and application authorization.
Outline-backed action source is checked at compilation time with
\texttt{this} bound to the declared receiver. Host-language implementations constitute trusted
deployment code governed by the surrounding application policy. The base Entitir action boundary
accepts a receiver-qualified entity identifier and invokes \texttt{executeAction}; both host
controllers and the generic model-facing action invocation can delegate to this boundary. The QRC
integration adds a distinct route for query-derived continuations: its model-facing adapter
requires an accepted receiver-selecting query, derives authority from the typed denotation, and
invokes only the permit-bearing dispatch entry point. The routes share the same action registry and
action implementation, while the QRC route assigns query-derived continuations the stronger entry
condition evaluated in this work.

Receiver validation rejects missing fields, illegal edge continuations, post-terminal chains, and
wrong-receiver actions before repository access. QRC additionally rejects query-provenance
substitution, target amplification, action or effect escalation, principal substitution, relevant
contract or selected-row changes, argument substitution, payload alteration, and replay.
Application permission, policy, and impact checks remain composable gates before repository
mutation.

\subsection{Events and external integration}
\label{sec:events}

Built-in create, update, and delete operations apply a local repository mutation inside the
enclosing transaction and immediately publish a typed \texttt{EntitirEvent} to the in-process
bus; the mutation becomes visible to other connections only when the dispatch transaction commits.
Custom domain actions participate through an event-producing mutation path or explicit event
publication. Configured subscribers run synchronously at publication and can apply
deployment-specific external-dispatch policy; because they run before commit, a subscriber
exception fails the current target (Section~\ref{sec:qrc-impl}), and a subscriber that performs
an external call does so for a mutation that is not yet committed. Production integrations
therefore attach a transactional outbox at this boundary and perform external delivery, retry,
compensation, and recovery after commit; an outbox makes delivery at-least-once, so the
receiving side must be idempotent or deduplicate on the mutation's identity.

\section{Evaluation}
\label{sec:evaluation}

\subsection{Evaluation design}

\subsubsection{Research questions}

\textbf{RQ1} asks whether the compositional contract preserves concrete receiver extensions,
propagates genuine result-type changes, rejects invalid continuations, and follows the ontology
world loaded for the request. \textbf{RQ2} asks whether QRC enforces receiver, target, operation,
principal, argument, provenance, replay, and dependency conditions under controlled attacks and
repository-backed execution. \textbf{RQ3} asks which ontology and data changes preserve an
existing approval. It compares policies that bind selected rows, recheck membership,
validate recorded observations, or require the whole repository state to remain unchanged.
\textbf{RQ4} measures the issuance, dispatch, storage, and
recovery costs of realizing the contract.

Table~\ref{tab:evaluation-summary} summarizes the comparisons and their endpoints.
The controlled decision studies use independently specified expected outcomes.
Repository experiments check the resulting state as well as the dispatch decision;
timing studies measure issuance and dispatch under a specified action requirement.

\begin{table}[!htbp]
\centering
\caption{Evaluation protocols and outcomes}
\label{tab:evaluation-summary}
\footnotesize
\setlength{\tabcolsep}{3pt}
\begin{tabular}{c>{\raggedright\arraybackslash}p{3.8cm}>{\raggedright\arraybackslash}p{3.7cm}>{\raggedright\arraybackslash}p{3.8cm}}
\toprule
RQ & Dataset or suite & Comparison & Primary endpoint \\
\midrule
1 & Scope, composition, receiver copies and 45 revisions & Receiver scope, future-\texttt{this}, live world & Member legality and result types \\
2 & 960 cases per condition; 30,000 generated trials; 60 workflows & Authority ablations, QRC, verifier mutants & Decisions and repository states \\
1--2 & 24 attribution runs; 6 subset cases; 18 query/schema variants & Qualified, root-owner, coarse dependencies & Dependency sets and action-entry decisions \\
3 & 360 delayed-workflow trials & Selected rows, witnesses, recheck, whole state & Eligible completions and selection re-executions \\
3 & Northwind: 78 cases per method & JDBC, QRC witnesses, QRC recheck & G2 decisions, effects and replay \\
3 & 12 configurations; 216 cases per method & QRC and JDBC under two approval policies & Derived keys and policy decisions \\
3 & Repeated types: 128 decisions and 12 maintenance trials & Closure, path evidence, membership & Precision and projection maintenance \\
4 & Scaling, placement, and ledger history & QRC, direct execution, result handle & Latency, storage, recovery \\
4 & 13 configurations; 1,404 lifecycles; 12 state-change trials & G2 witnesses, recheck, handle & Latency, reads, payload, contention \\
4 & 18 configurations in five JVMs; 6,480 lifecycles; 60 recheck decisions & Witness, full/target recheck, handle & Process costs, query plans, exact state \\
3--4 & SQLite prototype: 32 cases, 12 timing configurations & Closure, typed paths, automatic SQL, target recheck & Evidence equality, policy enforcement, cost \\
3--4 & Java: 64 precision and 216 evolution decisions; SQLite: four catalog changes & Strict path and closure; typed and SQL bindings & Exact state, cost, binding and rejection stage \\
3 & Eight Java query pairs; 54 SQLite scenarios per route & Equal answers; current and former evidence sources & Paired entries, false admissions and refusals \\
3 & Existing approvals: 40 Java and 30 SQLite cases per route & Typed navigation and stable-view SQL & Entry and replay after a change \\
\bottomrule
\end{tabular}
\end{table}

\subsubsection{Compositional typing protocol (RQ1)}
\label{sec:typing-protocol}

The receiver study compares future-\texttt{this} with a generic-base return in 24 signature
tests and 32 composition tests. It also checks three receiver copies, the same 32 signatures
in strict TypeScript~5.9.3, 140 receiver-scope pairs, and 45 world revisions. Supplementary
Sections~S3 and~S18 give the protocols and outcomes by family.

\subsubsection{Query-derived capability protocol (RQ2--RQ4)}
\label{sec:cq-protocol}

The evaluation compares QRC with five authorization configurations: a global catalog,
a receiver-scoped catalog that checks whether an operation applies to the receiver type,
signed-resource authority, target-row version authority, and whole-world version authority. A
server-side result-handle design shares QRC's derived payload and state-dependent decision rule,
isolating authority derivation from carried versus stored placement. Finally, single-premise
verifier mutants test whether the scenario oracles detect defective enforcement. These layers
complement the protocol-level GraphQL/OData comparison in Table~\ref{tab:authority-position} and
the TypeScript language reference used by RQ1.

The primary matrix contains 48 semantic templates across HR/security, employee onboarding, and
power-grid recovery, with 20 parameterized instances per template and condition. Its 180 legal
and 780 adversarial cases vary receiver, query provenance, target set, principal, action,
arguments, effect, dependency state, authenticated payload, and replay status. All six conditions
receive identical inputs, and expected decisions are encoded independently from the verifier.
The signed-resource and target-row version configurations differ only in whether they
revalidate the invoked target rows, so their paired decisions isolate the effect of that check.

A fixed-seed generator supplies 30,000 additional trials whose expected decisions follow
from Eq.~\eqref{eq:qrc-auth}. Sixty SQLite workflows then check repository mutations and
replay. The backend, mutation-testing and recovery protocols appear in Supplementary
Sections~S1, S12 and S13. The Java cost studies use an Apple M1 Pro with 32~GB RAM
and JDK~21.0.10. Two integration checks compare SQL and stream denotations and replay
360 fixed-corpus plans against independently stored outcomes.

\subsection{Results}
\label{sec:results}

\subsubsection{RQ1: Retention, transformation, and executable continuation}
\label{sec:typing-results}

In the 24-case ablation, future-\texttt{this} satisfies all 24 expectations and generic-base
returns satisfy 16. All eight differences are collection-extension continuations. Both arms
retain ordinary generic element information on the controls, so this result isolates concrete
receiver retention rather than the availability of generics.

In the 32-program composition suite, the full engine meets all 32 oracles versus 18 for
generic-base returns. Nine differences concern concrete collection receivers and five concern
bindings, callbacks, or navigation. Both arms meet all nine result-evolution and eight
invalid-program oracles. Independent TypeScript signatures reproduce all four family totals
(Supplementary Material, Section~S18). Three receiver-copy programs preserve both source objects
and retain their extension methods. Receiver-scoped validation admits all 40 legal programs and
rejects all 100 cross-receiver substitutions; the global catalog admits both. The live checker
meets 45/45 revision expectations versus 15/45 for the frozen catalog.

\subsubsection{Cross-stage attribution and target-scoped boundaries}
\label{sec:cross-stage-ablation}

We isolate receiver-qualified dependency ownership with a repository-backed ablation that
carries one query through validation, selection, issuance, and mutation entry. \texttt{Employee},
\texttt{Device}, and \texttt{Department} each declare \texttt{status}. Two chained filters read
\texttt{Employee.age} and \texttt{e.device().status}; the final action marks one selected employee.
The dependency owner changes to \texttt{Device} inside the predicate while the result and action
receiver remain \texttt{Employee}. A direct root and an immutable top-level root alias are both
fully SQL-lowered. The analyzer suite separately covers block-local navigation aliases.

Three conditions share the denotation, action, arguments, nonce handling, dispatcher, and
repository. The qualified condition uses the service-generated footprint. A root-owner ablation
replaces the navigated \texttt{Device.status} schema and value dependencies with
\texttt{Employee.status} dependencies. A coarse ablation retains the correct footprint and
also binds the same-named field on an unrelated department and an unreached device. Ablated
payloads are authenticated after changing their dependency sets, isolating trusted derivation
from payload tampering. Each condition is exercised under no change, a change to the referenced
device, a change on the unrelated owner, and a change on the unreached same-type device.

The independent eligibility oracle reads the referenced device directly; post-state checks
verify the invoked and non-invoked employees. Across the two query forms, qualified derivation
makes all eight expected decisions. Misattribution admits both stale-device continuations;
coarse attribution unnecessarily rejects four eligible continuations
(Table~\ref{tab:cross-stage-ablation}). These 24 controlled runs establish the contribution of
receiver-qualified ownership to both stale-continuation rejection and eligible-continuation
preservation at the query-to-action boundary.

\begin{table}[!htbp]
\centering
\caption{Effect of receiver attribution on action entry}
\label{tab:cross-stage-ablation}
\small
\begin{tabular}{@{}lrrr@{}}
\toprule
Derivation & Correct / 8 & \shortstack{Ineligible\\accepted / 2} & \shortstack{Eligible\\rejected / 6} \\
\midrule
Receiver-qualified & 8 & 0 & 0 \\
Root-owner attribution & 6 & 2 & 0 \\
Coarse same-name binding & 4 & 0 & 4 \\
\bottomrule
\end{tabular}
\end{table}

\paragraph{Equal answers, different approvals}
\leavevmode\phantomsection\label{exp:java-equal-answers}%
Two Java queries selected the same item, one through its parent and the other through an
alternate relation. Each received the same strict action grant. We changed neither,
one, or both candidate endpoints from 80 to 90, leaving both selections' threshold of
70 satisfied. With distinct endpoints, changing the parent alone invalidated only the
parent-based approval; changing the alternate alone reversed the verdicts. With a shared
endpoint, the two approvals agreed in all four cases. All 16 dispatches across the eight
pairs matched the independent row oracle and the expected repository state (Supplementary
Section~S21). The source of the observation distinguished the approvals even though the
answer and the action were identical.

Six additional cases use a constant filter with two selected roots and no field-state
witnesses. The unchanged case, deletion of the non-target root, and a change to that root are
accepted for a singleton target; deletion of the target, including a deleted root in the target
set, and an empty target set are rejected before action entry. All six cases agree with the
target-scoped contract. The non-target deletion case demonstrates that Theorem~2 preserves
membership for the invoked subset $A$ while allowing changes to the remainder of $D$.

A separate query/schema-evolution check rebuilds the ontology under three declarations:
the original path, renamed edge and field members, and a changed edge-target type. Six
queries per declaration cover root fields, navigation, alpha-renaming, local aliases,
chained filters, and a second navigation hop. All 18 derived dependency sets exactly match
independently specified owner-qualified keys, including exclusion of unrelated same-named
fields. This checks that the compiler follows the loaded ontology and accepted expression
as their names, paths, and dependency owners change. The alias cases test extraction;
issuance additionally requires the SQL-lowering check in Section~\ref{sec:execution}.

\subsubsection{Capability enforcement and operational behavior (RQ2, RQ4)}
\label{sec:cq-results}

Across 48 templates in three domains, QRC admits all 180 legal continuations and rejects
all 780 adversarial cases. Binding the whole snapshot also blocks all 780 but rejects 60
legal changes to an unreferenced declaration. The signed-resource baseline rejects 600 of
the 780 adversarial cases; adding target-row freshness raises that count to 660.
The remaining misses concern query provenance and
referenced action, policy, or executor dependencies. Supplementary Section~S18 gives the
complete six-condition matrix. Each condition receives the same inputs and labels.

A repository check distinguishes schema from row freshness. In a query reading only employee
fields, changing an unselected employee or the selected employee's laptop leaves the grant valid.
Changing an unreferenced column on the
\emph{invoked} employee changes that row's fingerprint and is rejected, even though the
column has no separate query dependency. The full-row freshness rule accounts for this refusal; the query's field dependencies
remain unchanged.

The independently constructed property generator satisfies all 30,000 metamorphic oracles:
6,924 authority-preserving trials are accepted and 23,076 authority-changing trials are rejected
over randomized denotations and payloads.

Mutation analysis confirms that the pass rates discriminate QRC from defective verifiers.
Thirteen of 14 single-premise mutants are killed: ten by the matrix and three by targeted
lifetime, world-binding, and dependency-presence probes. The remaining mutant is semantically
equivalent for the tested encoding: it omits the argument type check, while equality of canonical
argument digests already implies equal values and types. Removing both the type and digest
premises is killed by all 120 argument-substitution cases. The analysis therefore identifies the
explicit type check as defense in depth and demonstrates the additional discrimination supplied
by the boundary probes.

Across 60 repository-backed capabilities in three domains, every legal dispatch changed only
the intended row using its bound argument. All stale-target, amplification, principal-substitution,
and replay attempts were rejected (Supplementary Material, Section~S18). The eight boundary
regression categories also meet their oracles, including six PostgreSQL cases. In the issuance
race, a rival committed between two reads; the recorded fingerprint matched the state selected
by the query. Dispatch rejected the stale capability without consuming its nonce.

The result-handle reference reproduces all 960 accept/reject decisions. It retains the same
derived authorization payload server-side and gives the client a 43-byte reference. This
comparison isolates storage placement; the equal-G2 lifecycle comparison below evaluates
verification strategies under enforced action requirements.

\subsubsection{RQ3: Business-level delayed continuations}
\label{sec:business-workflows}
\label{sec:footprint-experiment}

This study compares continuation policies against a shared, independent business oracle.
Three workflows use one employee--device--department ontology:
an approved batch requires unchanged selected employees; device recovery additionally requires
an unreturned assigned device; department recovery additionally requires an active department.
The shared data model isolates progressively stronger continuation requirements. Each workflow
is crossed with ten committed evolution traces, selected-set sizes
1, 3, and 5, and four policies, giving 360 trials in fresh SQLite worlds. The traces change
selected or unselected rows, predicate or unreferenced fields at each navigation level, insert a
new matching employee, assign the same value, or change and restore a predicate value.
All policies share the selected set, grants, stable dispatcher, action, and post-state checks.
The \emph{selected rows} policy uses the default G1 footprint. The \emph{field-sensitive witnesses} policy derives
$F_F$ from the accepted query and binds the referenced scalar fields and navigation foreign keys
on each reached object of the corresponding type. \emph{Query recheck} reruns the original
selection in the same write transaction and invokes the selected-row dispatcher only if all
original targets still qualify. This comparator tests business eligibility. In the performance
study of Section~\ref{sec:equal-g2}, the action descriptor requires G2 for all methods, and
the full and target-restricted recheck methods test membership inside the dispatcher.
\emph{Whole-state binding} compares snapshots of the actual contents of all rows in
the three tables. Consequently, a
same-value assignment and a change followed by restoration do not invalidate the content-based
policies. The independent oracle reads repository fields directly and requires unchanged
selected employees plus the workflow's device and department conditions. It does not call the
QRC verifier or reuse the query translator.

\begin{table}[!htbp]
\centering
\caption{Workflow outcomes by continuation policy}
\label{tab:business-workflows}
\footnotesize
\setlength{\tabcolsep}{2.8pt}
\begin{tabular}{@{}lrrrr@{}}
\toprule
Policy & \shortstack{Eligible\\completed\\(/72)} &
\shortstack{Ineligible\\completed\\(/18)} &
\shortstack{Eligible\\rejected\\(/72)} &
\shortstack{Selection\\re-executions\\(/90 trials)} \\
\midrule
Selected rows & 72 & 9 & 0 & 0 \\
Field-sensitive witnesses & 72 & 0 & 0 & 0 \\
Query recheck & 72 & 0 & 0 & 90 \\
Whole-state binding & 27 & 0 & 45 & 0 \\
\bottomrule
\end{tabular}
\end{table}

Witness validation and query rechecking make the same 90 decisions: both complete all
72 eligible continuations and reject all 18 ineligible ones (Table~\ref{tab:business-workflows}).
Witness validation compares the recorded dependencies without re-executing the selection;
query rechecking runs it once per trial. Whole-state binding completes only 27 eligible cases.
Changes to an unreferenced device value or department label remain admissible under witness
validation; changes to \texttt{returned}, \texttt{active}, or a traversed foreign key are rejected.
The query counts quantify avoided re-execution; Section~\ref{sec:equal-g2} measures lifecycle
latency including witness capture and validation reads. Selected-row binding is sufficient for all 27 eligible approved
batches, but executes three ineligible device-recovery and six ineligible department-recovery
continuations. These nine outcomes identify workflows whose business requirement calls for G2
in addition to historical-selection authority.

Every dispatched action satisfies its target and non-target post-state checks. New matching
employees leave the original batch valid under both witness binding and target-membership
recheck, consistently with their G2 target-membership contract. Table~\ref{tab:business-workflows}
reports outcomes for the prescribed evolution traces; complete-current-answer equality is the
separate G3 requirement defined in Section~\ref{sec:scope}.

\subsubsection{Independent SQL and query/schema-evolution studies (RQ3)}
\label{sec:northwind}

The Northwind case study uses a public sample database~\cite{microsoft-northwind}:
77 products, 29 suppliers, and eight categories. Three inventory-review selections start with
active products whose stock is at or below the reorder level, add a UK-supplier condition,
and then add the Beverages category condition. The successive filters narrow the selection
from 18 products to four and then to one. The source values define the selections; a constructed review action
sets an added marker on approved targets. Thirteen continuation scenarios cross each query
with first-target and complete-result invocation, giving 78 cases per method. They vary
target and non-target rows, relevant and irrelevant linked fields, restored values, new
matches, target expansion, principal, operation, and empty targets.

The independent comparator uses JDBC, a server-side saved result, complete target-row
snapshots, and a separately written SQL query. It checks the saved principal, operation,
nonempty target subset, target-row equality, and current membership, then performs the update
inside \texttt{BEGIN IMMEDIATE}. It invokes neither the QRC analyzer nor its verifier,
SQL translator, or dispatcher. QRC uses either derived witnesses or protected rechecking,
with G2 declared on the same review action. A Java oracle over complete before/after states
scores decisions and exact mutations independently of both implementations.

All three methods complete the 42 eligible cases and reject the 36 ineligible cases
(Table~\ref{tab:northwind}). Every accepted invocation changes exactly its approved targets,
and all 42 immediate replays per method are rejected. The production authority compiler is
unchanged for this new schema. As the query adds supplier and category conditions, its
automatically derived schema footprint grows from three to five to seven keys; its concrete
field-witness closure contains 54, 18, and seven keys for the respective result sets.
The methods reached the same G2 decisions on these scenarios by different means:
the JDBC controller re-executed its application SQL, while functional witnesses validated
the query-derived closure without repeating the selection.

\begin{table}[!htbp]
\centering
\caption{Northwind action outcomes under G2}
\label{tab:northwind}
\small
\setlength{\tabcolsep}{4pt}
\begin{tabular}{@{}lrrr@{}}
\toprule
Method & \shortstack{Eligible\\completed} & \shortstack{Ineligible\\rejected} & \shortstack{Replays\\rejected} \\
\midrule
JDBC saved result & 42/42 & 36/36 & 42/42 \\
QRC query recheck & 42/42 & 36/36 & 42/42 \\
QRC functional witnesses & 42/42 & 36/36 & 42/42 \\
\bottomrule
\end{tabular}
\end{table}

\paragraph{End-to-end derivation and evidence policy}
\label{sec:contract-evolution}
This study carries query/schema adaptation through issuance and action execution in two
additional domains: purchase approval through supplier and certificate relations, and instrument
approval through calibration and standard relations. Each domain has its original schema,
a renamed edge/field version, and a changed edge-target type. One- and two-hop selecting
queries give twelve query/schema configurations. Each schema is loaded before issuance;
committed data changes then occur between approval and dispatch. The production compiler and
dispatcher are unchanged across configurations.

Each query selects three roots. First-target and complete-result invocations are crossed with
nine changes: no change, a changed target row, loss or preservation of a target's predicate,
loss or preservation of another selected root's predicate, an unread linked-field change,
an unreached-row change, and restoration of an observed value. Four methods yield 864 decisions
in fresh SQLite worlds. QRC uses either functional witnesses or query rechecking. An independent
JDBC controller stores the approved result and target rows, then implements the same two
policies using supplied selection SQL and, for recorded evidence, an explicit observation
projection. It invokes no ontology analyzer or QRC verifier. Each schema version supplies
the same selection semantics to all four methods.

An independent Java oracle checks current predicate membership and target-row equality.
A second oracle additionally compares the query-observed values and links for all three
approved roots. These implement, respectively, current-membership and recorded-evidence
approval. Predeclared scenario verdicts also check both executable oracles. All twelve issued
query/schema configurations produce the exact expected owner-qualified keys. Every method
matches its policy oracle on all 216 continuations, changes exactly the authorized targets,
and rejects every accepted invocation's replay.

\begin{table}[!htbp]
\centering
\caption{Continuation outcomes by approval contract}
\label{tab:evolution-contracts}
\small
\setlength{\tabcolsep}{4pt}
\begin{tabular}{@{}lrrr@{}}
\toprule
Approval policy & \shortstack{QRC\\completed} & \shortstack{JDBC\\completed} &
\shortstack{Additional refusals vs.\\membership policy} \\
\midrule
Current membership & 156 & 156 & 0 \\
Recorded evidence & 96 & 96 & 60 \\
\bottomrule
\end{tabular}
\end{table}

Table~\ref{tab:evolution-contracts} shows when binding the evidence changes the decision.
Of the 156 continuations that met the membership policy, 60 differed
from the approved evidence. In 36, an observation for an invoked root changed while its
predicate remained true. In the other 24, only the first root was invoked and the change
concerned another approved root: its predicate remained true in 12 cases and became false
in 12. Thus even an eligible target can require renewed approval when the evidence policy
covers the original batch. QRC derived that footprint from the ontology query; the JDBC
evidence controller used an explicit projection. Both reached the same decisions.

\paragraph{Footprint precision and projection maintenance}
\label{sec:footprint-precision}
In the procurement and laboratory queries, each step reaches a distinct receiver type,
so the path and closure footprints coincide. A repeated type can make the closure
include fields or links beyond the path traversed by the query.
A self-referential \texttt{Item} ontology supports one- and two-hop parent queries and a
two-branch query whose endpoints are shared or distinct. Each configuration contains six rows
and selects one root. Eight committed changes vary an observed value, a closure-only value
or link, an unreferenced column, an unreached row, or a restored value, alongside the unchanged
control. Four methods produce 128 first decisions: QRC witnesses, QRC rechecking, JDBC
membership, and JDBC path-evidence validation. Independently declared path keys, closure keys,
predicate decisions, and complete row states define the oracles.

\begin{table}[!htbp]
\centering
\caption{Evidence precision for repeated receiver types}
\label{tab:footprint-precision}
\small
\setlength{\tabcolsep}{4pt}
\begin{tabular}{@{}lrrr@{}}
\toprule
Configuration & Path keys & Closure keys & \shortstack{Additional path-eligible\\refusals / 8} \\
\midrule
One-hop parent & 3 & 12 & 2 \\
Two-hop parent & 4 & 12 & 2 \\
Shared branch endpoint & 4 & 16 & 2 \\
Distinct branch endpoints & 5 & 20 & 2 \\
\bottomrule
\end{tabular}
\end{table}

The compiler produces every expected closure key and covers every path key
(Table~\ref{tab:footprint-precision}). Per 32 cases, both membership methods accept 28,
JDBC path-evidence validation accepts 24, and QRC witnesses accept 16. Each matches its
specified policy oracle; all accepted actions have exact post-states, and all 96 accepted
invocations reject replay. Of the 24 cases accepted by path-specific validation, witnesses
reject eight: in each of the four configurations, one change affects only a scalar value beyond
the query path and another affects only a foreign key beyond it. Both changes preserve the
observed path and target membership but alter the recorded closure.

The corresponding adaptation audit supplies the JDBC controller with selecting SQL and a
separate observation projection, whereas QRC receives the ontology query and derives its
evidence. Both general controllers remain unchanged. Across the four configurations, three
distinct selecting forms require three explicit SQL observation projections and no separately
supplied QRC projection. A maintenance ablation then lengthens a navigation or adds a distinct
branch while omitting the newly read scalar from the SQL observation projection. The selection
itself is updated correctly. Two adaptations, unchanged/changed-evidence conditions, and three
methods give twelve trials. All methods accept the unchanged controls; the incomplete projection
admits both changed-evidence continuations, while the updated SQL projection and QRC reject both.
The two changed-evidence cases locate the synchronization cost: a controller with an explicit
projection needs an edit to that projection as well as to the selecting query. Automatic
derivation needs only the revised selection.

\subsubsection{Enforced requirements and equal-G2 performance (RQ4)}
\label{sec:equal-g2}

The implementation declares G2 on the evaluated action descriptor and uses the same
issuance service, dispatcher, selected-row checks, permissions, argument binding, durable nonce
ledger, and receiver-confined update for every method. \emph{Witness} derives functional
field-state evidence; \emph{Recheck} evaluates the authenticated selection inside dispatch;
\emph{Target-recheck} restricts that evaluation to the invoked identifiers; \emph{Handle}
stores the signed Witness payload durably and resolves it through the same dispatcher.
Handle includes its forced payload write, read, and removal.

For the row-local fragment $\mathcal{Q}_F$, predicate evaluation is independent for each
root, so $A\subseteq[\![Q]\!]_{S'}$ can be checked by evaluating the predicate on $A$.
The target-restricted comparator uses a primary-key \texttt{id IN (A)} condition and
independently written correlated SQL predicates for the benchmark's three query shapes.
It retains the signed query, target-row checks, and G2 descriptor. This experimental
adapter activates after functional-fragment classification, verifies the recognized query
shape, and leaves issuance unchanged. Rank, bounds, and global predicates retain full-query
rechecking.
Thirty paired correctness scenarios span three depths, one or three targets, and unchanged,
changed-target, eligibility-losing, eligibility-preserving, and unrelated changes. Both
recheck methods meet all 30 decision and exact-state oracles and reject every accepted
capability's replay. Query-plan inspection records a root scan for full rechecking and an
integer-primary-key search for target-restricted rechecking, whose navigation also uses
linked-row primary-key lookups. The benchmark creates no secondary predicate indexes.

Requirement regressions cover profile downgrade, descriptor replacement, cached and suspended
action entry, immutable argument binding, metadata recovery, and exact SQLite/PostgreSQL
post-states. Rejected invocations produce no protected-action entry. Twelve paired state trials
separate G2 from evidence equality: both methods reject membership loss and changed target rows,
and accept unrelated-row changes and new matches. An observed navigated value may change while
still satisfying an inequality; Witness then rejects the changed evidence, while Recheck
establishes membership from current values. The minimum guarantee is shared even when their
admissible continuation sets differ.

The initial 13-configuration sweep timed 1,404 complete lifecycles on file-backed
SQLite with a forced-write nonce ledger. With 1,000 selected roots and depth one,
Witness made 4,001 dispatch read calls, versus 1,002 for full rechecking: it captured
the whole approved result even for one target. Under eight-client contention, median
monitor-blocked time was 421.0~ms for Witness and 137.5~ms for Recheck. Supplementary
Sections~S12--S13 separate the cost of evidence reads from serialization and ledger
storage; the replicated comparison below adds an indexed, target-restricted baseline.

\paragraph{Independent-process cost surface}
The replicated study crosses candidate count $N\in\{10{,}000,100{,}000\}$, selected roots
$|D|\in\{10,100,1000\}$, and navigation depth $h\in\{0,1,3\}$, with one target and one
client. Five separate JVMs execute all 18 configurations. Each method/configuration has
three rounds of six measured lifecycles and three warm-ups per round; method order rotates
across rounds and processes. Four methods yield 6,480 measured lifecycles and 3,240 excluded
warm-ups. Each point in Fig.~\ref{fig:g2-surface} is the median of five paired ratios of
process-level lifecycle medians; bars show their observed minimum and maximum.
Values below one favor Witness.

In the first five-JVM series, Witness had a lower median lifecycle cost than full rechecking in
5 of 18 configurations. Target-restricted rechecking had a lower median in all 18
(Witness/Target-recheck: 1.066--5.292). A second five-JVM series favored Witness
over full rechecking in 5 configurations and favored target-restricted rechecking in 17.
The exception was a depth-zero query with ten selected roots among 100,000 candidates
(ratio 0.996). The optimized comparator removes the repeated root scan and
evaluates navigated predicates through indexed lookups on the invoked target.
Witness costs instead grow with the captured closure of the complete approved result.
These measurements distinguish two uses of the authority contract: derived fingerprints
preserve the recorded query observations without repeating the predicate, while indexed
rechecking provides a current-membership test for the evaluated row-local queries.
Both retain the same descriptor-enforced G2 requirement.
All methods share authority compilation and dispatch, isolating the cost of evidence
mechanisms within the same authorization contract.

\begin{figure}[!htbp]
\centering
\begin{tikzpicture}
\begin{groupplot}[
group style={group size=2 by 2,horizontal sep=1.1cm,vertical sep=1.6cm},
width=0.46\linewidth,height=5.2cm,
xmode=log,log basis x=10,xtick={10,100,1000},
xticklabels={10,100,1000},xlabel={Selected roots $|D|$},
ymin=0.5,ymax=5.7,ymajorgrids=true,grid style={gray!20},
tick label style={font=\small},label style={font=\small},
title style={font=\small},legend style={font=\small,draw=none},legend columns=3,
]
\nextgroupplot[title={$N=10{,}000$},ylabel={Witness / Full recheck},xlabel={}]
\addplot[gray,densely dotted,forget plot] coordinates {(10,1) (1000,1)};
\addplot+[black,solid,mark=*,thick,error bars/.cd,y dir=both,y explicit]
table[x=d,y=r,y error minus=lo,y error plus=hi,row sep=\\] {
d r lo hi\\
10 0.9936 0.0650 0.0177\\
100 1.3215 0.0890 0.0393\\
1000 1.7892 0.0825 0.0394\\
};
\addplot+[blue!65!black,dashed,mark=square*,thick,error bars/.cd,y dir=both,y explicit]
table[x=d,y=r,y error minus=lo,y error plus=hi,row sep=\\] {
d r lo hi\\
10 1.0958 0.0668 0.1193\\
100 1.8506 0.1909 0.0231\\
1000 3.3141 0.0368 0.0181\\
};
\addplot+[black,dashdotted,mark=triangle*,thick,error bars/.cd,y dir=both,y explicit]
table[x=d,y=r,y error minus=lo,y error plus=hi,row sep=\\] {
d r lo hi\\
10 1.0239 0.0086 0.0448\\
100 2.4013 0.1262 0.0422\\
1000 4.8042 0.0870 0.1618\\
};
\nextgroupplot[title={$N=100{,}000$},xlabel={}]
\addplot[gray,densely dotted,forget plot] coordinates {(10,1) (1000,1)};
\addplot+[black,solid,mark=*,thick,error bars/.cd,y dir=both,y explicit]
table[x=d,y=r,y error minus=lo,y error plus=hi,row sep=\\] {
d r lo hi\\
10 0.7918 0.0438 0.0318\\
100 0.9933 0.0372 0.0170\\
1000 1.6322 0.0050 0.0216\\
};
\addplot+[blue!65!black,dashed,mark=square*,thick,error bars/.cd,y dir=both,y explicit]
table[x=d,y=r,y error minus=lo,y error plus=hi,row sep=\\] {
d r lo hi\\
10 0.8057 0.0522 0.0319\\
100 1.3004 0.0368 0.0559\\
1000 2.9370 0.0182 0.0330\\
};
\addplot+[black,dashdotted,mark=triangle*,thick,error bars/.cd,y dir=both,y explicit]
table[x=d,y=r,y error minus=lo,y error plus=hi,row sep=\\] {
d r lo hi\\
10 0.8469 0.0467 0.0499\\
100 1.6545 0.0553 0.0590\\
1000 4.2580 0.1008 0.0263\\
};
\nextgroupplot[title={$N=10{,}000$},ylabel={Witness / Target recheck}]
\addplot[gray,densely dotted,forget plot] coordinates {(10,1) (1000,1)};
\addplot+[black,solid,mark=*,thick,error bars/.cd,y dir=both,y explicit]
table[x=d,y=r,y error minus=lo,y error plus=hi,row sep=\\] {
d r lo hi\\
10 1.0893 0.2153 0.1090\\
100 1.3943 0.0166 0.0575\\
1000 1.9692 0.0720 0.0293\\
};
\addplot+[blue!65!black,dashed,mark=square*,thick,error bars/.cd,y dir=both,y explicit]
table[x=d,y=r,y error minus=lo,y error plus=hi,row sep=\\] {
d r lo hi\\
10 1.1184 0.0883 0.1602\\
100 1.8530 0.0033 0.0384\\
1000 3.5973 0.0560 0.0806\\
};
\addplot+[black,dashdotted,mark=triangle*,thick,error bars/.cd,y dir=both,y explicit]
table[x=d,y=r,y error minus=lo,y error plus=hi,row sep=\\] {
d r lo hi\\
10 1.1890 0.0739 0.0215\\
100 2.4239 0.0305 0.3511\\
1000 5.2917 0.1840 0.0487\\
};
\nextgroupplot[title={$N=100{,}000$}]
\addplot[gray,densely dotted,forget plot] coordinates {(10,1) (1000,1)};
\addplot+[black,solid,mark=*,thick,error bars/.cd,y dir=both,y explicit]
table[x=d,y=r,y error minus=lo,y error plus=hi,row sep=\\] {
d r lo hi\\
10 1.0663 0.1168 0.0932\\
100 1.2389 0.0665 0.0759\\
1000 1.8913 0.0253 0.0267\\
};
\addplot+[blue!65!black,dashed,mark=square*,thick,error bars/.cd,y dir=both,y explicit]
table[x=d,y=r,y error minus=lo,y error plus=hi,row sep=\\] {
d r lo hi\\
10 1.1244 0.1114 0.0324\\
100 1.6712 0.0064 0.0383\\
1000 3.3926 0.0430 0.0735\\
};
\addplot+[black,dashdotted,mark=triangle*,thick,error bars/.cd,y dir=both,y explicit]
table[x=d,y=r,y error minus=lo,y error plus=hi,row sep=\\] {
d r lo hi\\
10 1.1446 0.0044 0.0700\\
100 2.1773 0.0772 0.0263\\
1000 4.9483 0.0129 0.0719\\
};
\legend{$h=0$,$h=1$,$h=3$}
\end{groupplot}
\end{tikzpicture}
\caption{Witness cost relative to query rechecking}
\label{fig:g2-surface}
\end{figure}

\subsubsection{Action-level evidence and two automatic derivations (RQ3--RQ4)}
\label{sec:prototype}

We built an independent SQLite research prototype to separate two questions: can an action
enforce an evidence policy across issuers, and does that policy require a type closure? A
typed-path extractor consumes the ontology filter. A second extractor consumes only the
selecting SQL and the trusted relational mapping: it adds each join's foreign key and each
predicate column together with its owner's row identifier. Neither receives an observation
projection. Both use the same action, grants, credential format, transaction, and ledger.
This SQL extractor is distinct from both the Java QRC runtime and the explicit-projection
JDBC controller above. Table~\ref{tab:prototype} pairs the prototype admission counts
with the cost of a split-branch query over 128 selected roots. The Java integration
is measured separately below.

For the four query/data configurations in Table~\ref{tab:footprint-precision}, we reran eight changes
per configuration and checked each decision and full post-state against the previously specified
oracle. Both automatic methods recovered the same physical keys and admitted 24 of 32
continuations; closure admitted 16. The eight recovered cases comprise four off-path scalar
changes and four off-path link changes. Target rechecking admitted four further cases in which
an observed value had changed but still met the threshold. All 92 successful dispatches
across methods rejected replay. Applying two changed selections to both extractors required
no separate projection edits. Across 51 policy, boundary, and adaptation checks, a membership
issuer could not issue a strict grant; a descriptor upgrade retired outstanding grants; and
a membership-only permit failed the protected entry predicate. Null links, cyclic data, shared
endpoints, and binder or SQL-alias renaming were checked separately (Supplementary Section~S19).

\begin{table}[!htbp]
\centering
\caption{SQLite prototype decisions and costs}
\label{tab:prototype}
\small
\setlength{\tabcolsep}{3pt}
\begin{tabular}{@{}lrrrr@{}}
\toprule
Method & Admitted$^{a}$ & Field keys & \shortstack{SELECTs\\issue/dispatch} & Lifecycle (ms) \\
\midrule
Type closure & 16/32 & 2,560 & 643/644 & 7.71 \\
Typed paths & 24/32 & 640 & 387/388 & 3.80 \\
Automatic SQL & 24/32 & 640 & 4/388 & 2.87 \\
Target recheck & 28/32 & 0 & 3/6 & 0.74 \\
\bottomrule
\end{tabular}

{\footnotesize $^{a}$Decisions span all 32 single-root scenarios; the other columns describe
the 128-root workload with one invoked target.}
\end{table}

The cost study crossed those four configurations with three selected-set sizes in five fresh processes
(1,920 measured lifecycles). All four methods in Table~\ref{tab:prototype} enforce G2:
each invoked target must still satisfy the selecting query at batch entry. Only the first
three---type closure, typed paths, and automatic SQL---also require the recorded observations
to match their issue-time values at that point. On the split-branch workload, typed paths reduce the number of
field keys from 2,560 to 640. SQL derives those same keys in a set-oriented projection,
requiring four issue-time SELECT statements versus 387 for per-row path traversal.
Supplementary Section~S19 reports the full
cost surface, timing protocol, and two-connection writer waits.

\paragraph{Integrated Java enforcement}
We repeated the precision experiment through Java's normal issuer, dependency snapshotter,
dispatcher, and protected action, with G2 and recorded evidence required for both profiles.
The 64 decisions reproduce the distinction above: finite paths admitted 24/32 and closure
16/32. Every observed key set and resulting database state matched its independent oracle.
Across the twelve procurement and laboratory configurations, the Java path issuer attributed
each field and edge to its expected owner. Its 216 dispatches matched the evidence oracle
(96 admissions, 120 refusals); every admitted invocation rejected replay.
On PostgreSQL, a changed shared score blocked a two-target batch before nonce consumption.
Restoring the score allowed the batch to run, while a second connection's update to the
shared endpoint completed after the transaction. A separate reopen test recovered the
strict descriptor and executed an authorized action through the path profile.

A paired Java/SQLite microbenchmark fixed the action, evidence policy, fresh file ledger,
and single invoked target. With one selected root, median lifecycle time was 8.72~ms for
finite paths and 10.38~ms for closure; with 32 selected roots it was 16.12 and 33.32~ms,
respectively (median of five JVM medians). The split-branch query required five path keys
versus twenty closure keys per root. These are within-Java comparisons under equal policy;
the automatic SQL extractor in Table~\ref{tab:prototype} was timed in Python.
Supplementary Section~S20 gives the Java and model-evolution protocols.

\paragraph{Interface changes in the shared-mapping SQLite study}
The Java runs above resolve fields and links against each loaded schema. In Java,
scalar members use their declared storage names (Table~\ref{tab:dependency-map}).
To isolate the effect of changing a logical name independently of a physical column,
we used two separate SQLite front ends with an explicit object--table, edge--key,
and field--column catalog. Both share the catalog, action descriptor, grant and
server-side handle. The action approves a purchase; the evidence is read through
its supplier and certificate. Each change in Table~\ref{tab:paired-evolution}
starts independently from this configuration.

\begin{table}[!htbp]
\centering
\caption{Selection edits and stale bindings in SQLite}
\label{tab:paired-evolution}
\small
\begin{tabular}{@{}>{\raggedright\arraybackslash}p{3.7cm}>{\raggedright\arraybackslash}p{2.5cm}>{\raggedright\arraybackslash}p{2.6cm}>{\raggedright\arraybackslash}p{3.3cm}@{}}
\toprule
Change & Typed selection & SQL selection & Old binding or grant \\
\midrule
Logical field rename & Edit field & Unchanged & Typed binding refused \\
Relation target table & Unchanged & Edit join & SQL binding refused \\
Physical column rename & Unchanged & Edit column & SQL binding refused \\
Action descriptor revision & Unchanged & Unchanged & Both old grants refused at dispatch \\
\bottomrule
\end{tabular}
\end{table}

In the retargeting case the old SQL still executed and returned the expected purchase ID:
only checking its join against the mapped target exposed the stale owner. After adaptation,
both routes recovered the same identities and observations and made the same action decisions.
Neither needed a separate observation projection. The difference was which names the selection
depended on: physical SQL insulated it from a logical rename, while typed navigation insulated
it from the two storage-facing changes. A SQL application can likewise expose stable names
through views. Counting selection edits alone does not capture this division of responsibility:
the mapping or view must still identify the current evidence owner.

\paragraph{Same answer, different evidence source}
We populated current and retired inventory sources with the same 77 Northwind stock counts.
The catalog designated the current source; the migration and subsequent changes were
controlled. Approvals were issued \emph{after} the catalog transition. All routes
received the same grant, issue-time IDs and stock values. We varied
the two sources independently, changing the stock of either the invoked product or another
approved product. The oracle compared issue-time observations for the approved result with
the current inventory; the common dispatcher applied each admitted action to one product.

\begin{table}[!htbp]
\centering
\caption{Action decisions after an inventory source change}
\label{tab:owner-intervention}
\small
\begin{tabular}{@{}lrrr@{}}
\toprule
Evidence binding & Correct & False entry & False refusal \\
\midrule
Typed navigation & 54 & 0 & 0 \\
SQL with updated join & 54 & 0 & 0 \\
SQL through current view & 54 & 0 & 0 \\
Retained former binding (control) & 30 & 12 & 12 \\
\bottomrule
\end{tabular}
\end{table}

Across 54 paired scenarios, each of the four routes admitted 18 actions. The routes
disagreed on which actions to admit (Table~\ref{tab:owner-intervention}): the former
binding missed changes to the current stock and refused actions after changes confined
to the retired source. A stable SQL view
over the current mapping preserved both the application query and the correct decisions.
The owner used to capture evidence must implement the relation used to approve the action.
Supplementary Section~S21 gives the interventions and the binding-stage checks.

\subsubsection{Validity of existing approvals after ontology changes (RQ3)}
\label{sec:old-approval}

To test the validity of an existing approval, we issued a credential, applied one
controlled change, and invoked the action with the original credential. In each of two domains, procurement and laboratory
inspection, a one-hop and a two-hop selection initially approved the same three objects.
The invocation targeted the first. Our decision rule, specified by transition and path
depth before scoring, preserves an approval if the edit leaves its recorded sources,
observations, target state and grant intact. Changing a referenced source invalidates
the approval even if the replacement contains identical IDs and values.

We compared the Java typed-path compiler with a trusted SQL adapter that read through a
stable view and resolved the installed view's projected columns and joins against the
same ontology. Both used the Java grant projector, signer, dependency verifier, nonce
ledger and action implementation. Each ran on its own copy of the same fixture.
As a read-only check, the post-change selections returned the three issue-time identifiers
in every scenario. When the second-hop field was renamed, this check used its new name
in the two-hop typed query and its updated mapping in the SQL view. Each route presented
the original credential to the dispatcher; neither used the check result as authority.

\begin{table}[!htbp]
\centering
\caption{Approval validity after ontology changes}
\label{tab:old-approval}
\small
\begin{tabular}{@{}p{7.2cm}cc@{}}
\toprule
Transition & One hop & Two hops \\
\midrule
Identical reload; declaration or action outside the footprint; change to an unselected endpoint row & Retain & Retain \\
First relation retargeted; granted action revised; bound policy version changed & Block & Block \\
Second relation retargeted; its scalar member renamed; observed value changed on another approved second-hop endpoint & Retain & Block \\
\midrule
Original approvals retained per route (two domains) & 14/20 & 8/20 \\
\bottomrule
\end{tabular}
\end{table}

Each route matched all 40 expected decisions: 22 original approvals entered once and
18 did not enter or consume the nonce. The six path-depth pairs in the third row of
Table~\ref{tab:old-approval} isolate the distinction: a change at the second hop
mattered only to the approval that used it. In particular, copying the related rows
to a different ontology type preserved both answers but not the identity of the declared
source. A missing declared field stopped Java fingerprint capture before action entry.

A physical rename need not change the declared source. We tested this separately in
a SQLite mapping adapter using the Northwind stock observations. A trusted migration
transaction recorded the old and new column for each managed rename under the same
logical owner and field. The adapter followed an unbroken record of those renames and
then compared the original observations; matching names or values alone was
insufficient. Typed navigation and stable-view SQL each retained 12 of 30 original
approvals and blocked 18. An unrelated rename preserved approval, as did two successive
recorded renames of the observed column. A missing step, a change of owner, or a change
to the recorded stock still blocked entry. This mapping facility belongs to the SQLite
adapter; the Java runtime binds scalar fields to their declared storage names.
Supplementary Section~S22 gives the transition schedules and decision counts.

\subsubsection{Ledger recovery and retained-history cost (RQ4)}
\label{sec:recovery-results}

Concurrent consumers admitted one of 32 attempts on the same nonce. Median consumption
time rose from 3.01~ms with an empty ledger to 28.97~ms at 100,000 retained records.
The file protocol, crash cases and complete history series appear in Supplementary
Section~S13. The lifecycle comparisons use matched fresh-history schedules.

\subsubsection{Integration evidence}

The surrounding execution pipeline supplies two further consistency checks. First, all 49
designed read compositions and all 900 scaling pairs agree with independently specified
denotations across their applicable execution modes. Second, of all 360 fixed-corpus plans, 266
satisfy the independent semantic oracle and preserve their expected outcomes, including 137
writes to existing receivers dispatched through QRC. Of the 94 semantically inexact plans,
90 did not request execution. Receiver validation blocked two of the four execution
requests, and the live-state check blocked the other two. None of the 94 plans produced
a mutation event.

\section{Discussion}
\label{sec:discussion}

\subsection{Findings and implications}

Authority compilation draws on three sources: the type at each access identifies the
declaration it reads, evaluation supplies the targets, and policy supplies the permitted
action and arguments. Misattributing a navigated member to the root admitted stale continuations;
binding every same-named member rejected eligible ones. Receiver-qualified attribution resolved
both errors. Carrying the resulting contract or storing it behind a handle gave the same
decisions in the controlled placement comparison.

The action requirement determines which comparison matters. For current membership alone,
indexed target rechecking had a lower median lifecycle cost than closure validation in
18 of 18 original configurations and 17 of 18 in the repetition. For approvals that also bind query observations, the prototype compares
two automatic derivations under that stronger requirement. Both follow a changed selection
without a separate projection edit, and both admit the same eight off-path changes that
invalidate the conservative closure. SQL obtains the keys from a join; typed navigation identifies their owners along the path.

The paired results distinguish the evidence contract from its derivation.
SQL over a maintained view derives the evidence owner from the installed join and
catalog; VirtualSet resolves the member on its receiver while traversing the validated
query. In Table~\ref{tab:old-approval}, both derivations feed the same Java issuer
and verifier. Their agreement shows that differently expressed selections can enforce
the same contract. The architectural difference is which binding must track a
retargeted relation: the SQL view or the ontology declaration used by receiver-flow analysis.

\subsection{Operational composition}
\label{sec:safeguards}

Controller policy determines which operations a request may exercise. If that decision
depends on external state, the provider must reflect any withdrawal of the issued grant
in a recorded dependency (Section~\ref{sec:trust}). In the Java regression in
Supplementary Section~S20, lowering an approval limit left the selected object and its
observations unchanged but invalidated the grant before action entry.

Built-in operations use per-target savepoints and publish synchronous repository events
before commit. A subscriber exception rolls back that target. External delivery requires
an application-level outbox and idempotent consumers. The shared ledger prevents repeated
consumption; registry and world monitors stabilize one process's configuration. Independently
changing registries require coordinated versions and locking for either carried capabilities
or stored handles.

\subsection{Applicability and evaluation coverage}
\label{sec:validity}

The contract separates historical authority (G1), current target membership (G2), and
complete current-answer equality (G3). For $\mathcal{Q}_F$, derived functional witnesses
establish G2; protected re-evaluation supplies an alternative. Target-restricted rechecking
uses root-local predicates, whereas ranked and global selections require their full
selection semantics. Aggregation and receiver-changing projection additionally require a
target-denotation contract before capability issuance. Explicit operator transitions and
denotation contracts provide the basis for extending authority compilation.

The evidence combines controlled ontology changes, independently specified eligibility
oracles, verifier mutants, exact repository post-states, and complete-corpus replay.
Northwind supplies a public schema and sample values. The procurement and laboratory studies
exercise complete issuance--dispatch lifecycles under rebuilt schemas and controlled changes.
The 22 retained approvals out of 40 Java scenarios and 12 out of 30 SQLite storage
scenarios per front end count specified treatments, not a sample of production edits.
The experiments identify which bindings change and whether the original approvals survive;
they do not measure developer time or the frequency of each change in practice.
The repeated-type study uses designed changes on and off the syntactic path. Its eight
closure-only refusals measure precision on those changes; their frequency in practice depends
on operational change histories. The Python/SQLite prototype isolates footprint construction
using a mapped self-relation, conjunctive filters and one protected action. Its execution metrics include SQL statement counts and the time a second connection waits to acquire a write transaction.
The Java study tests the QRC action entry, PostgreSQL metadata recovery and a concurrent
writer on a shared dependency. The paired Java timing study holds the strict evidence policy
fixed; the Python comparison additionally reports membership-only rechecking as a separate policy.
The separate SQLite studies exercise explicit schema mappings and a simple projected SQL
view. The stock values are public sample data; the migration and intervention frequencies
are experimental choices. Recorded evidence covers observations made by the query.
Other facts relevant to approval require their own trusted policy dependencies.

The PostgreSQL~18.1 regressions exercise transaction isolation, row locking, rollback,
and recovery of protected action descriptors. End-to-end tests on SQLite and PostgreSQL
check stored values and query expressions against the resulting updates; separate tests
cover protected-call composition. Performance evidence comprises a shared-world
concurrency sweep and five-process SQLite measurements. Multi-process configuration
coordination, an indexed transactional ledger, and an outbox for external delivery are
deployment extensions to the same authority contract.

\section{Conclusion}
\label{sec:conclusion}

VirtualSet connects receiver typing, dependency ownership, and trusted grants in a contract
for acting on query results. Target confinement and current membership concern the invoked
objects at batch entry; preserving particular observations is a further policy decision.
The Java runtime enforces both policy coordinates from issuance through protected action entry.
It can validate a conservative functional closure or record finite paths: a query about a
direct manager retains the link to that manager and the field it reads, without binding the
manager's manager. An approval can thus survive unrelated change without losing its basis.

When a second relation was retargeted to equal-valued records, only approvals that had
traversed it lost their basis. A maintained SQL view derived the same distinction under
the common verifier. VirtualSet traces each access in the validated query to its owning
receiver. The resulting capability tells the dispatcher which observations must still
hold before the authorized action begins.

\section*{CRediT authorship contribution statement}

\textbf{Qunhui Zhang:} Conceptualization, Methodology, Software, Validation, Formal analysis,
Investigation, Data curation, Writing--original draft, Visualization.

\section*{Funding}

This research received no specific grant from funding agencies in the public, commercial, or
not-for-profit sectors.

\section*{Declaration of competing interest}

The author declares no known competing financial interests or personal
relationships that could have appeared to influence the work reported in this paper.

\FloatBarrier


\begingroup
\small
\begin{thebibliography}{99}

\expandafter\ifx\csname url\endcsname\relax
  \def\url#1{\texttt{#1}}\fi
\expandafter\ifx\csname urlprefix\endcsname\relax\def\urlprefix{URL }\fi
\expandafter\ifx\csname href\endcsname\relax
  \def\href#1#2{#2} \def\path#1{#1}\fi

\bibitem{gruber-ontology-1993}
T.~R. Gruber, A translation approach to portable ontology specifications,
Knowledge Acquisition 5~(2) (1993) 199--220. doi:10.1006/knac.1993.1008.

\bibitem{poggi-obda-2008}
A.~Poggi, D.~Lembo, D.~Calvanese, G.~De Giacomo, M.~Lenzerini, R.~Rosati,
Linking data to ontologies, Journal on Data Semantics 10 (2008) 133--173.
doi:10.1007/978-3-540-77688-8\_5.

\bibitem{calvanese-ontop-2017}
D.~Calvanese, B.~Cogrel, S.~Komla-Ebri, R.~Kontchakov, D.~Lanti, M.~Rezk,
M.~Rodriguez-Muro, G.~Xiao, Ontop: Answering SPARQL queries over relational
databases, Semantic Web 8~(3) (2017) 471--487. doi:10.3233/SW-160217.

\bibitem{yu-spider-2018}
T.~Yu, R.~Zhang, K.~Yang, M.~Yasunaga, D.~Wang, Z.~Li, J.~Ma, I.~Li, Q.~Yao, S.~Roman, Z.~Zhang, D.~Radev, Spider: A large-scale human-labeled dataset for complex and cross-domain semantic parsing and text-to-{SQL} task, in: Proceedings of EMNLP, 2018, pp. 3911--3921. doi:10.18653/v1/D18-1425.

\bibitem{li-bird-2023}
J.~Li, B.~Hui, G.~Qu, J.~Yang, B.~Li, B.~Li, B.~Wang, B.~Qin, R.~Cao, R.~Geng, N.~Huo, X.~Zhou, C.~Ma, G.~Li, K.~C. Chang, F.~Huang, R.~Cheng, Y.~Li, Can {LLM} already serve as a database interface? A {BIg} bench for large-scale database grounded text-to-{SQL}s, Advances in Neural Information Processing Systems 36 (2023). doi:10.52202/075280-1835.

\bibitem{scholak-picard-2021}
T.~Scholak, N.~Schucher, D.~Bahdanau, {PICARD}: Parsing incrementally for constrained auto-regressive decoding from language models, in: Proceedings of EMNLP, 2021, pp. 9895--9901. doi:10.18653/v1/2021.emnlp-main.779.

\bibitem{pourreza-2023-dinsql}
M.~Pourreza, D.~Rafiei, {DIN-SQL}: Decomposed in-context learning of text-to-{SQL} with self-correction, in: Advances in Neural Information Processing Systems (NeurIPS), 2023. doi:10.52202/075280-1577.

\bibitem{gao-dailsql-2024}
D.~Gao, H.~Wang, Y.~Li, X.~Sun, Y.~Qian, B.~Ding, J.~Zhou, Text-to-{SQL} empowered by large language models: A benchmark evaluation, Proceedings of the VLDB Endowment 17~(5) (2024) 1132--1145. doi:10.14778/3641204.3641221.

\bibitem{wang-mac-sql}
B.~Wang, C.~Ren, J.~Yang, X.~Liang, J.~Bai, L.~Chai, Z.~Yan, Q.-W. Zhang, D.~Yin, X.~Sun, Z.~Li, {MAC-SQL}: A multi-agent collaborative framework for text-to-{SQL}, in: Proceedings of COLING, 2025, pp. 540--557.

\bibitem{gao-xiyansql-2024}
Y.~Gao, Y.~Liu, X.~Li, X.~Shi, Y.~Zhu, Y.~Wang, S.~Li, W.~Li, Y.~Hong, Z.~Luo, J.~Gao, L.~Mou, Y.~Li, A preview of {XiYan-SQL}: A multi-generator ensemble framework for text-to-{SQL}, arXiv preprint arXiv:2411.08599 (2024).

\bibitem{li-api-bank-2023}
M.~Li, Y.~Zhao, B.~Yu, F.~Song, H.~Li, H.~Yu, Z.~Li, F.~Huang, Y.~Li, {API}-Bank: A comprehensive benchmark for tool-augmented {LLM}s, in: Proceedings of EMNLP, 2023, pp. 3102--3116. doi:10.18653/v1/2023.emnlp-main.187.

\bibitem{ruan-toolemu-2024}
Y.~Ruan, H.~Dong, A.~Wang, S.~Pitis, Y.~Zhou, J.~Ba, Y.~Dubois, C.~J. Maddison, T.~Hashimoto, Identifying the risks of {LM} agents with an {LM}-emulated sandbox, in: International Conference on Learning Representations, 2024.

\bibitem{yao-taubench-2025}
S.~Yao, N.~Shinn, P.~Razavi, K.~Narasimhan, $\tau$-bench: A benchmark for tool-agent-user interaction in real-world domains, in: International Conference on Learning Representations, 2025.

\bibitem{lu-toolsandbox-2025}
J.~Lu, T.~Holleis, Y.~Zhang, B.~Aumayer, F.~Nan, H.~Bai, S.~Ma, S.~Ma, M.~Li, G.~Yin, Z.~Wang, R.~Pang, ToolSandbox: A stateful, conversational, interactive evaluation benchmark for {LLM} tool use capabilities, in: Findings of NAACL, 2025, pp. 1160--1183. doi:10.18653/v1/2025.findings-naacl.65.

\bibitem{dli-paper}
Q.~Zhang, Generic Constraints Projection: Four-Dimensional Type Inference for Dynamic Languages, arXiv preprint arXiv:2607.19693v4 [cs.PL] (2026). doi:10.48550/arXiv.2607.19693.

\bibitem{w3c-sparql-2013}
S.~Harris, A.~Seaborne, SPARQL 1.1 Query Language, W3C Recommendation, 2013.
\url{https://www.w3.org/TR/sparql11-query/} (accessed 29 July 2026).

\bibitem{opencypher-2026}
openCypher Implementers Group, openCypher: Open standards for graph query languages, 2026.
\url{https://opencypher.org/} (accessed 29 July 2026).

\bibitem{tinkerpop-gremlin-2026}
Apache Software Foundation, Apache TinkerPop Gremlin, 2026.
\url{https://tinkerpop.apache.org/gremlin.html} (accessed 29 July 2026).

\bibitem{typeql-2026}
Vaticle, TypeQL language reference, 2026.
\url{https://typedb.com/docs/typeql-reference/} (accessed 29 July 2026).

\bibitem{google-lookml-2026}
Google Cloud, What is LookML?, 2026.
\url{https://docs.cloud.google.com/looker/docs/what-is-lookml} (accessed 28 July 2026).

\bibitem{malloy-docs-2026}
Malloy Project, Malloy documentation, 2026.
\url{https://docs.malloydata.dev/} (accessed 28 July 2026).

\bibitem{cube-docs-2026}
Cube Dev, Cube documentation: Introduction, 2026.
\url{https://docs.cube.dev/docs/introduction} (accessed 28 July 2026).

\bibitem{dbt-docs-2026}
dbt Labs, dbt Developer Hub, 2026.
\url{https://docs.getdbt.com/} (accessed 28 July 2026).

\bibitem{zhou-ontology-tools-2026}
X.~Zhou, P.~Bulter, C.~Yang, S.~D. Rihm, T.~Angkanaporn, J.~Akroyd,
S.~Mosbach, M.~Kraft, Ontology-to-tools compilation for executable semantic constraint enforcement in LLM agents,
arXiv preprint arXiv:2602.03439 (2026). doi:10.48550/arXiv.2602.03439.

\bibitem{palantir-ontology-2026}
Palantir Technologies, Ontology overview, 2026.
\url{https://www.palantir.com/docs/foundry/ontology/overview} (accessed 28 July 2026).

\bibitem{palantir-actions-2026}
Palantir Technologies, Action types: Rules, 2026.
\url{https://www.palantir.com/docs/foundry/action-types/rules} (accessed 28 July 2026).

\bibitem{degiacomo-obda-updates-2017}
G.~De Giacomo, D.~Lembo, X.~Oriol, D.~F. Savo, E.~Teniente, Practical update management in
ontology-based data access, in: The Semantic Web -- ISWC 2017, Lecture Notes in Computer Science
10587, Springer, 2017, pp. 225--242. doi:10.1007/978-3-319-68288-4\_14.

\bibitem{botoeva-inseparability-2016}
E.~Botoeva, C.~Lutz, V.~Ryzhikov, F.~Wolter, M.~Zakharyaschev, Query-based entailment and inseparability for {ALC} ontologies, in: Proceedings of the Twenty-Fifth International Joint Conference on Artificial Intelligence, 2016, pp.~1001--1006.

\bibitem{canning-fbound-1989}
P.~Canning, W.~Cook, W.~Hill, J.~C. Mitchell, W.~Olthoff,
F-bounded quantification for object-oriented programming, in: Proceedings of the Fourth
International Conference on Functional Programming Languages and Computer Architecture, 1989,
pp. 273--280. doi:10.1145/99370.99392.

\bibitem{bruce-binary-1995}
K.~B. Bruce, L.~Cardelli, G.~Castagna, The Hopkins Objects Group, G.~T. Leavens, B.~Pierce,
On binary methods, Theory and Practice of Object Systems 1~(3) (1995) 221--242.
doi:10.1002/j.1096-9942.1995.tb00019.x.

\bibitem{saito-self-2009}
C.~Saito, A.~Igarashi, Self type constructors, in: Proceedings of the 24th ACM SIGPLAN Conference
on Object-Oriented Programming Systems, Languages, and Applications, 2009, pp. 263--282.
doi:10.1145/1640089.1640109.

\bibitem{java-stream-2026}
Oracle, Java Platform, Standard Edition 21 API Specification: Stream,
\url{https://docs.oracle.com/en/java/javase/21/docs/api/java.base/java/util/stream/Stream.html}
(accessed 20 September 2026).

\bibitem{typescript-this-2026}
Microsoft, TypeScript Handbook: Classes, this types,
\url{https://www.typescriptlang.org/docs/handbook/2/classes.html#this-types}
(accessed 20 September 2026).

\bibitem{scala-collections-2026}
Scala Center, Implementing Custom Collections (Scala 2.13),
\url{https://docs.scala-lang.org/overviews/core/custom-collections.html}
(accessed 20 September 2026).

\bibitem{graphql-spec-2021}
GraphQL Foundation, GraphQL Specification, October 2021 edition, 2021. \url{https://spec.graphql.org/October2021/} (accessed 28 July 2026).

\bibitem{meijer-linq-2006}
E.~Meijer, B.~Beckman, G.~M. Bierman, {LINQ}: Reconciling object, relations and {XML} in the {.NET} framework, in: Proceedings of the 2006 ACM SIGMOD International Conference on Management of Data, 2006, p. 706. doi:10.1145/1142473.1142552.

\bibitem{odata-csdl-2020}
OASIS, OData Common Schema Definition Language (CSDL) XML Representation Version 4.01,
OASIS Standard, 2020.
\url{https://docs.oasis-open.org/odata/odata-csdl-xml/v4.01/os/odata-csdl-xml-v4.01-os.html}
(accessed 24 August 2026).

\bibitem{odata-url-2020}
OASIS, OData Version 4.01, Part 2: URL Conventions, OASIS Standard, 2020.
\url{https://docs.oasis-open.org/odata/odata/v4.01/os/part2-url-conventions/odata-v4.01-os-part2-url-conventions.html}
(accessed 24 August 2026).

\bibitem{odata-protocol-2020}
OASIS, OData Version 4.01, Part 1: Protocol, OASIS Standard, 2020.
\url{https://docs.oasis-open.org/odata/odata/v4.01/os/part1-protocol/odata-v4.01-os-part1-protocol.html}
(accessed 15 September 2026).

\bibitem{dennis-vanhorn-1966}
J.~B. Dennis, E.~C. Van Horn, Programming semantics for multiprogrammed computations,
Communications of the ACM 9~(3) (1966) 143--155. doi:10.1145/365230.365252.

\bibitem{miller-robust-composition-2006}
M.~S. Miller, Robust Composition: Towards a Unified Approach to Access Control and Concurrency
Control, Ph.D. thesis, Johns Hopkins University, Baltimore, MD, 2006.

\bibitem{birgisson-macaroons-2014}
A.~Birgisson, J.~G. Politz, U.~Erlingsson, A.~Taly, M.~Vrable, M.~Lentczner,
Macaroons: Cookies with contextual caveats for decentralized authorization in the cloud,
in: Network and Distributed System Security Symposium, 2014.
doi:10.14722/ndss.2014.23212.

\bibitem{biscuit-2026}
Eclipse Biscuit Project, Biscuit: A bearer token with offline attenuation and decentralized
verification, specification, 2026. \url{https://doc.biscuitsec.org/reference/specifications}
(accessed 15 September 2026).

\bibitem{rfc9396-rar}
T.~Lodderstedt, J.~Richer, B.~Campbell, OAuth 2.0 Rich Authorization Requests,
RFC 9396, Internet Engineering Task Force, 2023. doi:10.17487/RFC9396.

\bibitem{park-sandhu-ucon-2004}
J.~Park, R.~Sandhu, The {UCON}$_{ABC}$ usage control model, ACM Transactions on Information and
System Security 7~(1) (2004) 128--174. doi:10.1145/984334.984339.

\bibitem{green-provenance-2007}
T.~J. Green, G.~Karvounarakis, V.~Tannen, Provenance semirings, in: Proceedings of the
26th ACM SIGMOD-SIGACT-SIGART Symposium on Principles of Database Systems, 2007, pp. 31--40.
doi:10.1145/1265530.1265535.

\bibitem{muller-sql-provenance-2018}
T.~M\"uller, B.~Dietrich, T.~Grust, You say `what', I hear `where' and `why':
(Mis-)interpreting SQL to derive fine-grained provenance, Proceedings of the VLDB
Endowment 11~(11) (2018) 1536--1549. doi:10.14778/3236187.3236204.

\bibitem{ahmad-dbtoaster-2012}
Y.~Ahmad, O.~Kennedy, C.~Koch, M.~Nikolic, DBToaster: Higher-order delta processing for
dynamic, frequently fresh views, Proceedings of the VLDB Endowment 5~(10) (2012) 968--979.
\url{https://www.vldb.org/pvldb/vol5/p968_yanifahmad_vldb2012.pdf}.

\bibitem{stonebraker-query-modification-1974}
M.~Stonebraker, E.~Wong, Access control in a relational data base management system
by query modification, in: Proceedings of the 1974 ACM Annual Conference, 1974,
pp. 180--186. doi:10.1145/800182.810400.

\bibitem{griffiths-wade-1976}
P.~P. Griffiths, B.~W. Wade, An authorization mechanism for a relational database
system, ACM Transactions on Database Systems 1~(3) (1976) 242--255.
doi:10.1145/320473.320482.

\bibitem{rizvi-fgac-2004}
S.~Rizvi, A.~Mendelzon, S.~Sudarshan, P.~Roy, Extending query rewriting techniques for
fine-grained access control, in: Proceedings of the 2004 ACM SIGMOD International Conference on
Management of Data, 2004, pp. 551--562. doi:10.1145/1007568.1007631.

\bibitem{nguyen-provenance-access-2012}
D.~Nguyen, J.~Park, R.~Sandhu, Dependency path patterns as the foundation of access
control in provenance-aware systems, in: 4th USENIX Workshop on the Theory and
Practice of Provenance (TaPP), 2012.

\bibitem{kung-robinson-occ-1981}
H.~T. Kung, J.~T. Robinson, On optimistic methods for concurrency control, ACM Transactions on
Database Systems 6~(2) (1981) 213--226. doi:10.1145/319566.319567.

\bibitem{zigmond-shilldb-2020}
E.~Zigmond, S.~Chong, C.~Dimoulas, S.~Moore, Fine-grained, language-based access control
for database-backed applications, The Art, Science, and Engineering of Programming
4~(2) (2020), Article~3. \url{https://doi.org/10.22152/programming-journal.org/2020/4/3}.

\bibitem{rfc7662-2015}
J.~Richer (Ed.), OAuth 2.0 Token Introspection, RFC 7662, IETF, 2015. doi:10.17487/RFC7662.

\bibitem{eckmann-hlr-sql-2026}
T.~Eckmann, M.~Urban, J.-M.~Bodensohn, C.~Binnig,
{HLR-SQL}: Human-like reasoning for Text-to-{SQL} with the human in the loop,
Information Systems 138 (2026) 102670. doi:10.1016/j.is.2025.102670.

\bibitem{bird-interact-2026}
BIRD Team, {BIRD-Interact}: Reimagining text-to-{SQL} evaluation via the lens of dynamic
interactions, in: International Conference on Learning Representations, 2026.
\url{https://bird-interact.github.io/} (accessed 29 July 2026).

\bibitem{poesia-synchromesh-2022}
G.~Poesia, O.~Polozov, V.~Le, A.~Tiwari, G.~Soares, C.~Meek, S.~Gulwani, Synchromesh: Reliable code generation from pre-trained language models, in: International Conference on Learning Representations, 2022.

\bibitem{beurerkellner-2023-lmql}
L.~Beurer-Kellner, M.~Fischer, M.~Vechev, Prompting is programming: A query language for large language models, Proceedings of the ACM on Programming Languages 7 (PLDI) (2023) Article 186. doi:10.1145/3591300.

\bibitem{willard-2023}
B.~T. Willard, R.~Louf, Efficient guided generation for large language models, arXiv preprint arXiv:2307.09702 (2023).

\bibitem{microsoft-typechat-2023}
Microsoft, TypeChat: Building natural language interfaces with types, 2023. \url{https://github.com/microsoft/TypeChat} (accessed 28 July 2026).

\bibitem{debenedetti-agentdojo-2024}
E.~Debenedetti, J.~Zhang, M.~Balunovi\'{c}, L.~Beurer-Kellner, M.~Fischer, F.~Tram\`{e}r,
{AgentDojo}: A dynamic environment to evaluate prompt injection attacks and defenses for {LLM}
agents, in: Advances in Neural Information Processing Systems 37, Datasets and Benchmarks Track,
2024.

\bibitem{debenedetti-camel-2025}
E.~Debenedetti, I.~Shumailov, T.~Fan, J.~Hayes, N.~Carlini, D.~Fabian, C.~Kern, C.~Shi,
A.~Terzis, F.~Tram\`{e}r, Defeating prompt injections by design, arXiv preprint arXiv:2503.18813v2
(2025). doi:10.48550/arXiv.2503.18813.

\bibitem{shi-progent-2025}
T.~Shi, J.~He, Z.~Wang, H.~Li, L.~Wu, W.~Guo, D.~Song, Progent: Securing {AI} agents with
privilege control, arXiv preprint arXiv:2504.11703v3 (2026). doi:10.48550/arXiv.2504.11703.

\bibitem{wang-authgraph-2026}
P. Wang, Y. Li, Y. Tian,
Aligning Provenance with Authorization: A Dual-Graph Defense for LLM Agents,
arXiv preprint arXiv:2605.26497 (2026).
\url{https://doi.org/10.48550/arXiv.2605.26497}.

\bibitem{li-pace-2026}
F. Li, Q. Wang, Y. Hu, K. Li, J. Liu, H. Wu, J. Zhou, D. Wang,
PACE: Provenance-Aware Capability Enforcement for Tool-Using LLM Agents,
arXiv preprint arXiv:2610.01349 (2026). \url{https://doi.org/10.48550/arXiv.2610.01349}.

\bibitem{xiong-securegard-2026}
G. Xiong, S. Wen, X. Chen, X. Zeng, Z. Fan, W. Ye, X. Cao, S.-J. Huang,
SecureGARD: Dynamic Ontology-Guided Governance for Secure Tool-Using LLM Agents,
arXiv preprint arXiv:2611.06682 (2026). \url{https://doi.org/10.48550/arXiv.2611.06682}.

\bibitem{saltzer-schroeder-1975}
J.~H. Saltzer, M.~D. Schroeder, The protection of information in computer systems, Proceedings of the IEEE 63~(9) (1975) 1278--1308. doi:10.1109/PROC.1975.9939.

\bibitem{sandhu-rbac-1996}
R.~S. Sandhu, E.~J. Coyne, H.~L. Feinstein, C.~E. Youman, Role-based access control models, Computer 29~(2) (1996) 38--47. doi:10.1109/2.485845.

\bibitem{gray-transactions-1981}
J.~Gray, The transaction concept: Virtues and limitations, in: Proceedings of the 7th International Conference on Very Large Data Bases, 1981, pp. 144--154.

\bibitem{garcia-molina-sagas-1987}
H.~Garcia-Molina, K.~Salem, Sagas, ACM SIGMOD Record 16~(3) (1987) 249--259. doi:10.1145/38714.38742.

\bibitem{microsoft-northwind}
Microsoft, Northwind and pubs sample databases for Microsoft SQL Server.
\url{https://github.com/microsoft/sql-server-samples/tree/master/samples/databases/northwind-pubs}.

\end{thebibliography}
\endgroup
\end{document}